\documentclass[aps,pra,showpacs,twocolumn]{revtex4-1}
\usepackage{graphicx,tabularx,booktabs,longtable,color,enumerate,hyperref}
\setcitestyle{numbers,square}

\begin{document}


\title{Assessing the impact of representational and contextual problem features \\on student use of right-hand rules}


\author{Mary Bridget Kustusch}
\affiliation{DePaul University, Department of Physics, 2219 North Kenmore Avenue Suite 211, Chicago, IL 60614, USA}
\email{mkustus1@depaul.edu}


\date{\today}

\begin{abstract}
Students in introductory physics struggle with vector algebra and these challenges are often associated with contextual and representational features of the problems. Performance on problems about cross product direction is particularly poor and some research suggests that this may be primarily due to misapplied right-hand rules. However, few studies  have had the resolution to explore student use of right-hand rules in detail. This study reviews literature in several disciplines, including spatial cognition, to identify ten contextual and representational problem features that are most likely to influence performance on problems requiring a right-hand rule. Two quantitative measures of performance (correctness and response time) and two qualitative measures (methods used and type of errors made) were used to explore the impact of these problem features on student performance. Quantitative results are consistent with expectations from the literature, but reveal that some features (such as the type of reasoning required and the physical awkwardness of using a right-hand rule) have a greater impact than others (such as whether the vectors are placed together or separate). Additional insight is gained by the qualitative analysis, including identifying sources of difficulty not previously discussed in the literature and revealing that the use of supplemental methods, such as physically rotating the paper, can mitigate errors associated with certain features.

\end{abstract}


\maketitle



\section{Introduction\label{intro}}
The use of physical mnemonics (\emph{e.g.} right-hand rules) to remember the direction of various physical quantities has a history dating from the late 1800's \cite{Greenslade:1980}. In the present day, most introductory physics sequences incorporate right-hand rules (RHRs) to some extent, typically in rotational dynamics and/or magnetism.  

The goal of these various RHRs is arguably to unload some of the mental processes required to deal with the 3-dimensional quantities related through vector cross products \footnote{\citet{Hestenes:Oersted} argues for eliminating the use of cross products. While there is merit to this approach, the ubiquitousness of cross products in the canonical introductory physics sequence warrants a better understanding of student use of cross products before making such a shift.}. Yet, there is some evidence, and rampant speculation, that RHRs bring their own set of challenges that may be associated with particular representational or contextual features of the problem \cite{Klatzky:2008,Scaife:2010,VanDeventer:thesis}. 

The current paper (based on the author's dissertation~\cite{Kustusch:dissertation}), explores the following question:
\begin{quote}
How do contextual and representational problem features impact student performance on problems requiring right-hand rules?
\end{quote}

The focus on contextual and representational dependence relies on two theoretical frameworks: resources and cognitive load theory. The resources framework  claims that knowledge is composed of networks of small resources that can be activated in various ways depending on the context~\cite{Hammer:2005}. Cognitive load theory claims that tasks with the same inherent difficulty (intrinsic cognitive load) may require more or less mental effort depending on the contextual and representational features of the task (extraneous and/or germane cognitive load)~\cite{Plass:2010}. In the light of these theories, helping students use RHRs effectively requires a better understanding of how their use of RHRs varies with context and representation.

In order to emphasize the use of RHRs, this paper focuses on those RHRs associated with the generation of magnetic fields from moving charges and magnetic force acting on moving charges. At this point in the introductory physics sequence, most students already have some familiarity with RHRs from rotational dynamics.

Section~\ref{literature} uses literature from several research domains to identify the features that are most likely to affect student performance and what we would expect that impact to be. Section~\ref{design} discusses the design of this study, focusing on operationalizing the features identified in Section~\ref{literature} and identifying appropriate measures of performance. Sections~\ref{quantitative} and \ref{qualitative} present quantitative and qualitative analyses of the data, respectively.
Finally, Section~\ref{conclusions} reviews the findings from both analyses and discusses the instruction implications of this research.

 \begin{figure*}
\begin{center}
\includegraphics[width=7.1in]{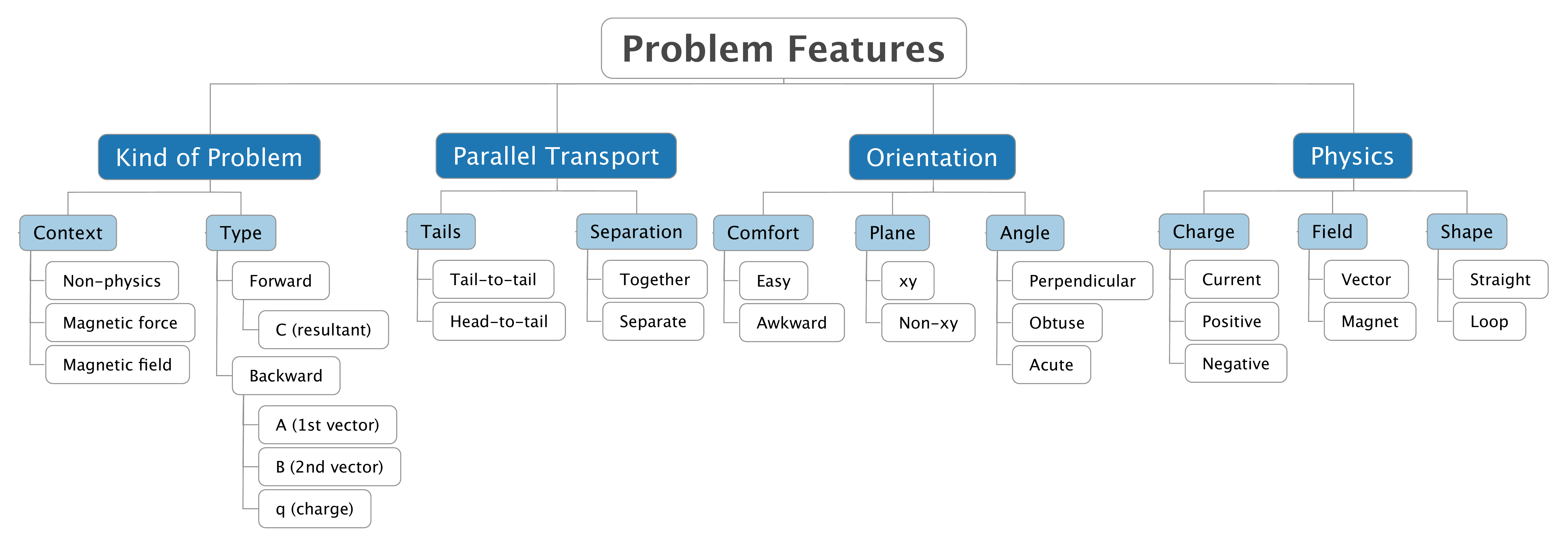}
\caption{Overview of the problem features identified from the literature and used in this study as predictors of performance. Individual features are in light blue (single quotations in text), with the variations beneath in white (italics in text). The features are grouped into larger categories (dark blue)) to facilitate discussion in the text. See Table~\ref{hypotheses} for the hypothesized impact of each feature on performance. }
\label{problem features map}
\end{center}
\end{figure*}

\section{Previous research relevant to the use of right-hand rules\label{literature}}
There are several domains of research that have implications for understanding student use of RHRs. This study draws primarily on the following areas: 
\begin{itemize}
\item conceptual understanding of magnetism (\ref{magnetism}),
\item understanding of vector algebra (\ref{vector}) , and
\item spatial cognition, specifically mental rotations and form perception (\ref{spatial}). 
\end{itemize}
Based on a synthesis of the research in these fields, this section identifies the contextual and representational features that are most likely to affect student performance on problems requiring the use of a RHR (Fig.~\ref{problem features map}) and what we  expect that impact to be (Table~\ref{hypotheses}). In the text, the problems features will be identified with single quotation marks and the variations with italics.

\begin{table}
\caption{Hypothesized impact of predictors on performance. All except the last three (which serve as control variables) are shown in light blue in Fig.~\ref{problem features map}. Angle, Charge, Shape, and Field were treated separately in the quantitative analysis due to the constraints of the data.}
\label{hypotheses}
\begin{ruledtabular}
\begin{tabular}{lp{2.5in}} 
\toprule
Predictor & Hypothesis\\ 

\hline
\\
Context & Participants will perform better on non-physics questions than on magnetic field or magnetic force questions and may perform better on magnetic force than on magnetic field questions.\\
\\
Type & Participants will perform better on questions that ask for the cross product result (forward reasoning) than on those that ask for one of the given vectors (backward reasoning).\\
\\
Tails & Participants will perform better on questions with vectors presented tail-to-tail than on those presented head-to-tail.\\
\\
Separation & Participants will perform better on questions where the vectors are together than on those separate in space.\\
\\
Comfort & Participants will perform better on questions with physically easy orientations than on those with awkward orientations.\\
\\
Plane & Participants will perform better on questions in the $xy$ plane than on those in a non-$xy$ plane ($xz$ or $yz$ plane).\\
\\
Angle & Participants will perform better on questions with perpendicular vectors than on those with acute or obtuse vectors.\\
\\
Charge & Participants will perform better on questions with positive charge or a current-carrying wire than on those with negative charge.\\
\\
Shape & Participants will use a different right-hand rule for a straight current-carrying wire and loop of current, but there will be no difference in performance.\\ 
\\
Field & Participants will perform better on questions where the magnetic field is represented by vectors than on those with magnets.\\
\\
\hline
\\
Order & The order of the questions will be positively correlated with performance due to practice effects. \\
\\
Experience & Participants with more experience (2nd group) will perform better than participants with less experience (1st group).\\
\\
Spatial & Spatial ability will be positively correlated with performance.\\
\\
 \end{tabular}
\end{ruledtabular}
\end{table}

\subsection{Magnetism}\label{magnetism}
There have been several attempts in recent years to develop and validate a conceptual survey of electromagnetic concepts analogous to the Force Concept Inventory (FCI) (e.g., \citet{Ding:2006,Maloney:2001}). Through these tools, as well as more qualitative methods,
the literature on student understanding of electromagnetism identifies several key areas where students struggle with the fundamental concepts of magnetism. 

One of the major sources of confusion is differentiating between electric and magnetic concepts~\cite{Guisasola:2002,Maloney:2001,Scaife:2010,Scaife:2011}. According to \citeauthor{Scaife:2010} \cite{Scaife:2010,Scaife:2011}, this confusion was  more common on magnetic force questions when the magnetic field was represented by magnetic poles than when it was represented with field lines (`Field'). The magnetic poles representation was also related to a sign error resulting from reversing the direction of the magnetic field due to those poles (south to north instead of north to south). They also identified two other sources of these non-random sign errors: ``a confusion in choice and execution of the several right-hand rules available'' and an unawareness of the non-commutative nature of the cross product. 

There is also evidence that students struggle more with electric field problem than with electric force problems~\cite{Garza:2012}, suggesting the possibility of a similar pattern with magnetic field  and force problems (`Context').

In addition to confusion between electric and magnetic field and forces, conventional wisdom also assumes that the need to reverse the direction of the cross product for a negative charge adds an additional layer of difficulty (`Charge'). Thus, we would expect that participants would perform better on problems with a positive charge or a current-carrying wire than on problems with a negative charge.

Also, there are many different RHRs that one can use, particularly for different current distributions, such as a straight wire or a loop of wire (`Shape'). One might expect the shape of the current distribution to affect the type of RHR used, while not necessarily affecting the accuracy of the response.

\subsection{Vector Algebra}\label{vector}
The research on student understanding of vector algebra has included studies on student conceptual understanding of vector properties \cite{Aguirre:1984} and the design and assessment of diagnostic tools to assess the level of incoming university students' ability to do vector operations~\cite{Knight:1995,Nguyen:2003,Barniol:2013}. 
More relevant to the current study are those that explore the sources of student difficulty with basic vector concepts and operations~\cite{Gagatsis:2001,Zavala:2009PERC} and those that compare performance on vector operations in different contexts~\cite{VanDeventer:2007,Garza:2012,Zavala:2012,Zavala:2013}, with different representations~\cite{Hawkins:2009,Barniol:2010,Scaife:2010,Heckler:2015}, and/or with different interventions~\cite{Flores:2004,Southey:2014}.

The majority of these studies focus on vector addition and subtraction using an arrow representation in a mechanics contexts (\emph{e.g.}, force, displacement, momentum,~\emph{etc.}). Zavala and colleagues \cite{Garza:2012,Zavala:2010,Zavala:2012,Barniol:2013,Zavala:2013} and \citeauthor{Scaife:2010} \cite{Scaife:2010,Scaife:2011} are have also studied vector multiplication, to some extent, and within the context of electromagnetism. However, only \citeauthor{Scaife:2010} \cite{Scaife:2010} had the resolution to actually observe the use of RHRs. 

There are clear indications from this research that performance on vector algebra problems is both representation-dependent and context-dependent. One common source of difficulty in graphical vector addition and subtraction is the need for parallel transport: moving the vectors in space while preserving length and direction \cite{Hawkins:2009,Nguyen:2003,Barniol:2010}. Many RHRs assume that the vectors are tail-to-tail. Thus, if the vectors are presented head-to-tail (`Tails') and/or separated in space (`Separation'), one would expect students to have more difficulty than if they are presented tail-to-tail and placed together.

In addition, students tend to ignore directional information and will often use an incorrect angle or trigonometric function to find components or calculate the magnitude of vector products. While using an inappropriate trigonometric function may partially be due to confusion between dot products and cross products, \citet{Barniol:2013} found the use of the correct trigonometric function was much higher on dot product problems than on cross product problems. For cross products, questions where the vectors are not at right angles (`Angle') are likely more challenging than those where the vectors are completely perpendicular to each other.

Several of the studies mentioned here examine performance on vector algebra problems in one or more physics context and compare it to performance in a non-physics (or math) context. In some cases, it appears that the physical situation may cue appropriate resources for students \cite{Zavala:2012}. However, when asked to find the direction from a cross product, there is evidence that students struggle more when the problem is in a physics context than when it is in a non-physics context \cite{VanDeventer:thesis} (`Context'). 

Finally, there is a clear consensus in this literature that vector subtraction is more challenging than vector addition for students (although the alignment of the vectors makes a difference) \cite{Heckler:2015}. Particularly with the arrow representation, vector subtraction is sometimes taught as if one were reasoning backward from vector addition (\emph{e.g.}, for $\vec{A}-\vec{B}$, what vector should be added to $\vec{B}$ in order to get $\vec{A}$?). Thus, we would expect that cross product problems that require reasoning backward (\emph{e.g.}, given the resultant and one initial vector, what is a possible direction for the other initial vector?) would be more challenging than those that require forward reasoning (`Type').

\subsection{Spatial Cognition}\label{spatial}
Since a vector cross product is inherently three-dimensional, dealing with cross products often requires spatial reasoning analogous to performing a mental rotation. The use of a physical RHR introduces a kinesthetic element to this reasoning, but it does not eliminate the spatial nature of the cognition required. There is a large body of literature dealing with spatial cognition that provides some indications of what factors might influence the use of a RHR. Due to the breadth of this field, this discussion is restricted to a few key studies that have a direct bearing on the use of RHRs. 

According to \citet{Klatzky:2008}, the primary cognitive issue in using a RHR is one of coordinating and aligning reference frames. They identify several  frames that must be aligned in order to appropriately use a RHR: those defined by the vectors, by the page, and by the hand relative to the body. From this perspective of coordinating reference frames, they also provide a task analysis for determining the direction of the cross product of two vectors $\vec{U} \times \vec{V}$:
\begin{quote}
``The performer must (a) create a reference frame in which $\vec{U}$ and  $\vec{V}$ lie on a common plane at an arbitrary location in self-defined space, (b) treat the hand as an external object within the vector-defined frame and map hand-centered coordinates into that frame so that the wrist lies at the vector intersection, the hand's cross-section is aligned with $\vec{U}$, and  $\vec{V}$ is closer to the palm than the back of the hand; (c) observe the direction of the thumb in space and align it with the third dimension in the vector-defined coordinate system.'' (p. 171)\cite{Klatzky:2008} (Note: this RHR is referred to as \emph{RHR:Type:Standard} in Table~\ref{method code definitions brief})
\end{quote}
\citeauthor{Klatzky:2008} state that the primary difficulties of this task are creating the frame from the vectors and getting one's hand into the appropriate position in that frame. While the use of the body may often facilitate cognition~\cite{Alibali:2005}, 
\citeauthor{Klatzky:2008} argue that the spatial reasoning required for the more intricate postures is correspondingly complex and these unusual positions could increase the cognitive difficulty instead of decrease it.

Other teachers and researchers have also suggested that difficulties with implementing a right hand rule are mainly kinesthetic: the accidental use of one's left hand~\cite{RHRPhoto:2002}; losing track by becoming so focused on the manipulation of the hand~\cite{Nguyen:2005}; the physical difficulty of aligning one's body in the proper orientation~\cite{Klatzky:2008}; and in some cases, the inability to use one's hand at all~\cite{VanDomelen:1999}. All of these authors assume that physical awkwardness contributes to difficulty with RHRs, although none provide experimental evidence to support this assumption.

Research on alignment has also identified a human sensitivity to orientation, usually in the contexts of form perception~\cite{Rock:1973,Rock:1983,Rock:1986};  imagined rotation, translation, and transformation~\cite{Klatzkyetal:1998,Shelton:2001}; and the kinesthetic aspects of spatial reasoning~\cite{Avraamides:2004,Flanders:1995}. This literature indicates that the ``top'' of an image, as well as the vertical direction, have a special place in human perception and in the ability to perform mental rotations. For example, \citeauthor{Pani:1993} and colleagues~\cite{Pani:1993,Pani:1994,Pani:1996} found that a vertical axis of rotation --- and to a lesser extent, the horizontal --- are the only directions where participants could consistently perform mental rotations accurately. \citet{Shelton:2001} also found that salient environmental properties (such as walls and furniture) strongly influenced  mental representation of the space and performance on spatial tasks. These results, along with the task analysis discussed above~\cite{Klatzky:2008}, suggest that the use of a RHR would be more difficult in situations when the vectors are not aligned with the local reference frame (e.g., vectors are at an angle and therefore not aligned with the axes of the paper) or when the axis of rotation is not vertical (e.g., one of the given vectors is not in $xy$-plane).

Despite these implications from spatial cognition research, \citet{Scaife:2010} did not see any measurable impact due to vector orientation while conducting individual interviews with students solving magnetic force problems. In order to resolve this difference in the literature, the current study attempts to separately address three different aspects of orientation: 
\begin{itemize}
\item  physical awkwardness (`Comfort');
\item  axis of rotation or plane of the vectors (`Plane'); 
\item  alignment with the local reference frame (`Angle'). 
\end{itemize}
The results show that by treating these issues separately, each aspect of orientation does have a measurable effect on performance that is consistent with the spatial cognition literature.

\subsection{Summary of previous research}
Drawing on three different research domains, this section has identified ten different problem features that are likely to impact performance. These features and their hypothesized impact on performance are summarized in Fig.~\ref{problem features map} and Table~\ref{hypotheses}, respectively. In order to facilitate discussion in the text, the features have been grouped into slightly larger categories, as shown in Fig.~\ref{problem features map}. 
\begin{itemize}
\item \emph{Kind of Question}---broad characteristics of the problem: the physics (or non-physics) context (`Context') and the type of reasoning (`Type'). 
\item \emph{Orientation}---features related to spatial cognition: physical awkwardness (`Comfort'), the plane of the given vectors (`Plane'), and the angle between the given vectors (`Angle'). 
\item \emph{Parallel Transport}---features indicating whether a vector will need to be moved: whether the vectors are head-to-tail or tail-to-tail (`Tails'), and whether they are together or separate (`Separation'). 
\item \emph{Physics}---features most closely related to the physical situation: the sign of the charge (`Charge'), how the magnetic field is represented (`Field'), and the shape of the current distribution (`Shape'). 
\end{itemize}
Also included in Table~\ref{hypotheses} are three control variables that will be discussed in the following section.

\section{Study Design\label{design}}
In order to assess the impact of the features identified in Section~\ref{literature} on student performance, we need to first operationalize how we will distinguish between the variations within a given feature, as well as how we will measure performance. After providing details about the study context, including the participants and tasks,
this section will discuss how the problem features and performance were operationalized in this study.

\subsection{Study context}\label{study context}

The author conducted individual semi-structured interviews with participants recruited from three sections~\footnote{Each section was taught by a different instructor, but all involved interactive lectures with studio-style labs taught by graduate students.} of a calculus-based introductory physics sequence, all using the third edition of \emph{Matter and Interactions} \citep{MI3:em}.  Fifteen students were interviewed within two weeks of finishing the unit on magnetic fields. After finishing the unit on magnetic force, an additional twelve students were interviewed. The two groups were distinguished from each other to control for any effect due to more exposure to the content (`Experience').

Every interview was recorded from two different perspectives (overhead and side view) and there were two major tasks given to participants. The first task was the Cube Comparison Test \cite{Ekstrom:1976}, which was used to provide a measure of individual spatial ability that could be controlled for in the quantitative analysis (`Spatial'). The second task was a think-aloud protocol where participants solved many cross product direction questions and the features in Fig.~\ref{problem features map} were varied across relevant problems. Given the large number of problems, there was the possibility of improvement over the course of the interview (a practice effect). To address this possibility, each participant received the problems in the same, partially randomized order~\footnote{There were a few unintentional exceptions to this discussed in \citet{Kustusch:dissertation}} and the quantitative analyses controlled for this order (`Order').

At the end of the interview, the interviewer asked every participant four follow-up questions:
\begin{enumerate}
\item Describe the right-hand rule(s) that you use.
\item How does the direction of $\vec{A} \times \vec{B}$ compare to the direction of $\vec{B} \times \vec{A}$?
\item On a scale of one to ten, what is your comfort level or confidence with using the right-hand rule?
\item Is there anything you found difficult about these problems that you have not already mentioned?
\end{enumerate}
 Although not analyzed in detail, responses to these questions were used to supplement the qualitative analysis. For example, responses to Question 1 were used as a starting point for descriptions of different types of right-hand rule and 
 Question 2 provided an explicit probe assessing whether the participant was aware of the non-commutative nature of the cross product.

\subsection{Operationalizing problem features}\label{problem features}
This section will provide examples of how the problem features in Fig.~\ref{problem features map} were varied in this study.

 \begin{figure}
\begin{center}
\includegraphics[width=3.4in]{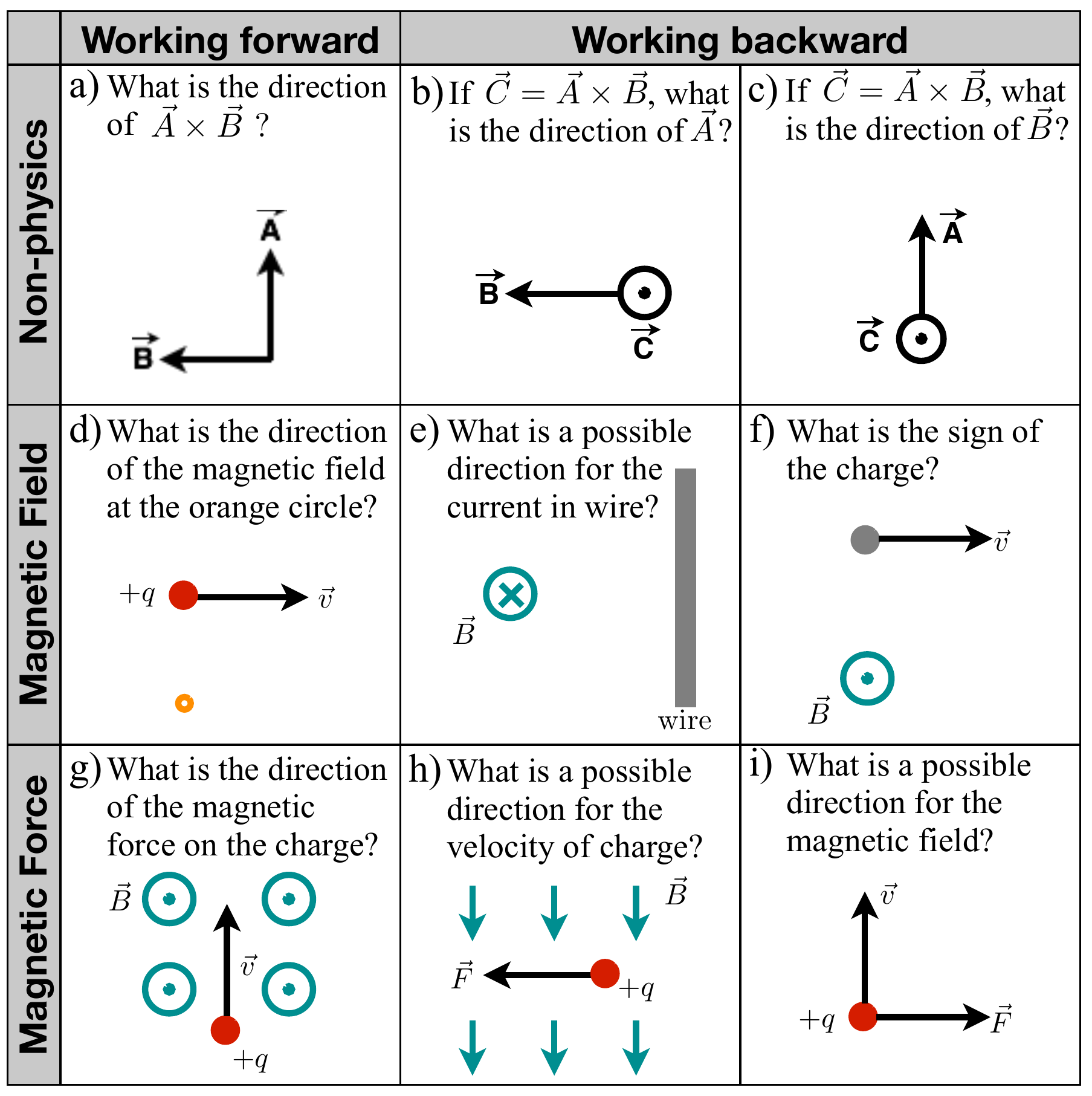}
\caption{Examples of \emph{Non-physics} (row 1), \emph{Magnetic Field} (row 2), and \emph{Magnetic Force} (row 3) questions where the problem involves working \emph{Forward} (column 1) and working \emph{Backward} (columns 2 and 3). Problem statements for rows 2 and 3 have been abbreviated for space.}
\label{context}
\end{center}
\end{figure}

\subsubsection{Kind of question: `Context' and `Type'}
 Questions were given in one of three contexts.
\begin{itemize}
\item \emph{Magnetic Field} questions involved the generation of magnetic field from moving charges. 
\item \emph{Magnetic Force} problems involved the force on moving charges due to an external magnetic field. 
\item \emph{Non-physics} problems were framed as $\vec{C}=\vec{A} \times \vec{B}$, with no explicit physics context. 
\end{itemize}
 
 In each context, there were questions that involved \emph{Forward} reasoning and questions that involved \emph{Backward} reasoning. Reasoning forward meant determining the cross product resultant vector from the given vectors (``What is the direction of $\vec{A} \times \vec{B}$?''). Reasoning backward questions provided the resultant vector ($\vec{C}$) and asked for one of the initial vectors (either $\vec{A}$ or $\vec{B}$). In physics contexts, a \emph{Backward} question might also ask for the sign of the charge or direction of current. Figure~\ref{context} shows examples of \emph{Forward} and \emph{Backward} problems for all contexts.
 
For logistical reasons, problems were presented to students in several sets based on `Context' and `Type.' Participants first answered the \emph{Magnetic Field} problems, with \emph{Forward} and \emph{Backward} questions intermixed. The second group of interviewees then answered \emph{Magnetic Force} questions, again with \emph{Forward} and \emph{Backward} questions intermixed \footnote{This group also answered two questions requiring a synthesis of both magnetic field and force concepts, but these questions were not included in the analysis}. After the physics problems were completed, both groups answered the \emph{Non-physics} questions, which were given as two sets of 25 \emph{Forward} questions and 
two sets of 10 \emph{Backward} questions.

\subsubsection{Parallel Transport: `Tails' and `Separation'}
\begin{figure}
\begin{center}
\includegraphics[scale=0.5]{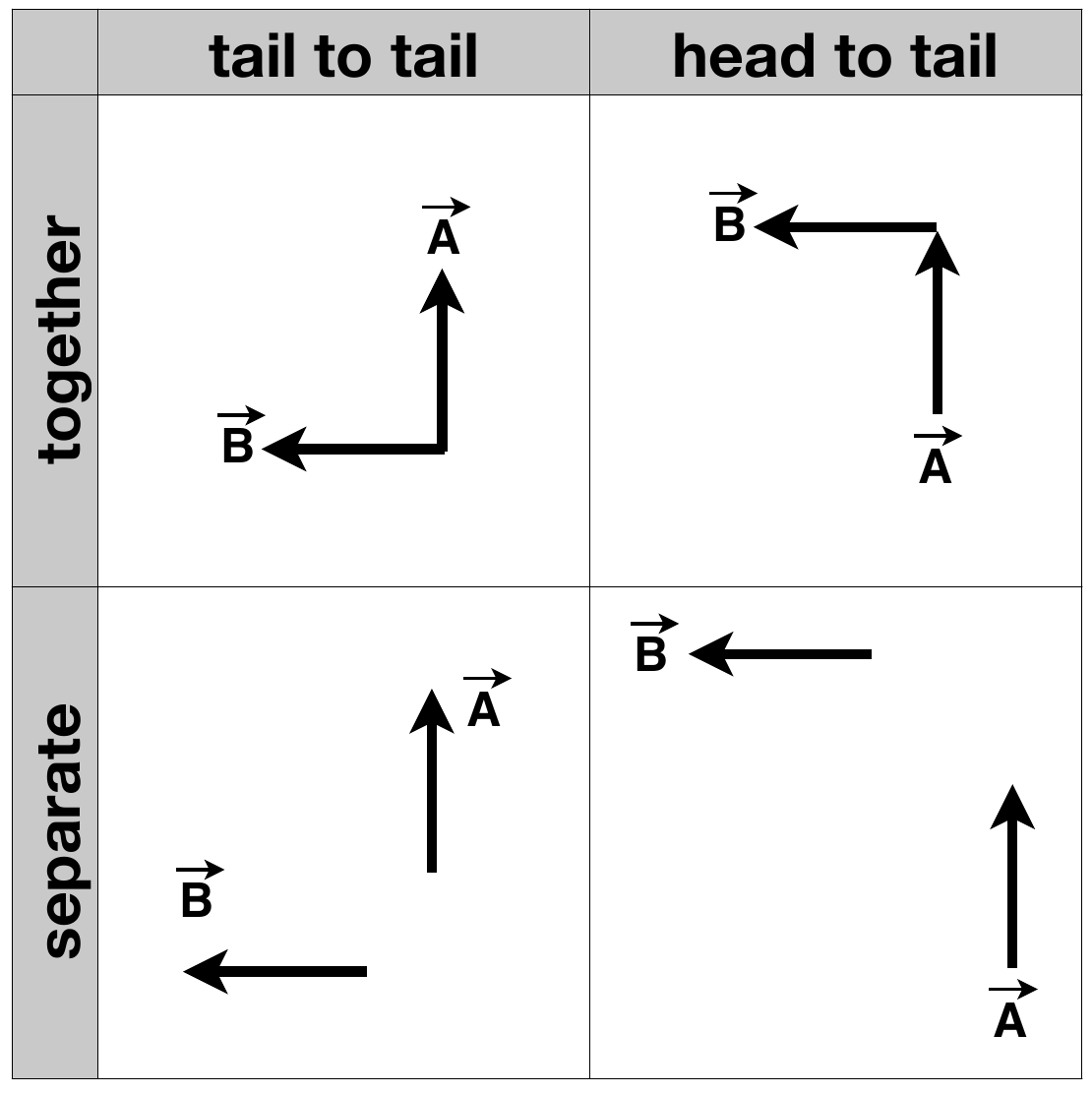}
\caption{The same orientation, $(+\hat{y})\times(-\hat{x})=+\hat{z}$,  with the vectors \emph{Head-to-tail}, \emph{Tail-to-tail}, \emph{Together}, and \emph{Separate}.}
\label{parallel transport}
\end{center}
\end{figure}

Problems were also varied in terms of the position of the vectors in space. This included whether the vectors were presented as \emph{Tail-to-tail} or \emph{Head-to-tail}, as well as whether they were \emph{Together} or \emph{Separate} in space. Figure~\ref{parallel transport} shows examples of the various possible combinations of these two features for one orientation. The non-physics problems included all of these combinations, whereas the physics problems included only those that were consistent with how physics problems are typically presented. Also, for \emph{Magnetic Field} problems, the position vector was not provided explicitly, only the observation location was indicated (see Fig.~\ref{context}.d for an example).

\subsubsection{Orientation: `Comfort', `Plane', and `Angle'}
\begin{figure}
\begin{center}
\includegraphics[width=3.25in]{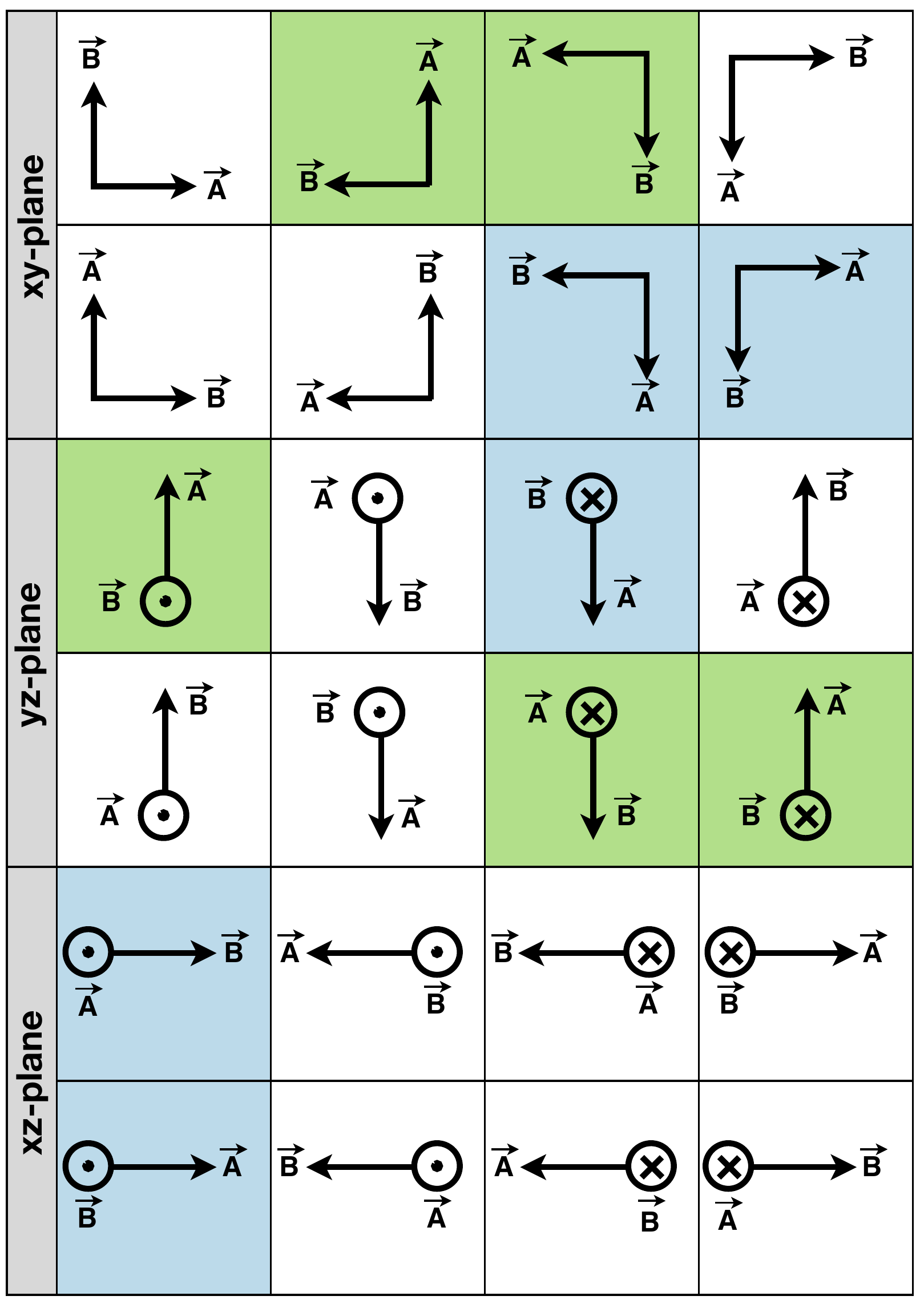}
\caption{All possible cross product combinations for on-axis, perpendicular vectors ($\vec{A}\times\vec{B}$). The green orientations were the five orientations with the lowest average ranking for physical awkwardness (\emph{Easy} orientations) and the blue orientations were those with the highest average ranking for physical awkwardness (\emph{Awkward} orientations).}
\label{physical discomfort}
\end{center}
\end{figure}

As discussed in Section~\ref{spatial} there are three aspects of the orientation of the vectors that were likely to contribute to difficulty: physical awkwardness (`Comfort'), the plane of the vectors (`Plane'), and the angle between the vectors (`Angle'). There are 24 different orientations of perpendicular, on-axis vectors in a 3-dimensional orthonormal coordinate space (Fig.~\ref{physical discomfort}). In order to restrict the possibilities and to quantify the level of physical awkwardness for various orientations, three physics graduate students rated their physical discomfort when using a right-hand rule to find the cross product direction for each of these 24 orientations. They rated them on a 5-point scale, with 1 being physically easy and 5 being physically awkward. The rankings for each orientation were averaged and the five orientations with the highest rank (\emph{Awkward} orientations) and the five orientations with the lowest rank (\emph{Easy} orientations) were used as a basis set for all of the interview problems.

It is important to note that whether an orientation is physically easy or physically awkward is dependent on both the method used and the errors made. Different RHRs require different hand positions and there are additional complications for physics questions, such as whether one flips for an electron at the beginning or end of the problem. There are also numerous errors that change the awkwardness of the problem, such as reversing the order of the cross product ($\vec{B} \times \vec{A}$ instead of $\vec{A} \times \vec{B}$). However, using these ten orientations allowed for a first-order comparison of the differences in performance on physically easy and physically awkward problems.

\begin{figure}
\begin{center}
\includegraphics[scale=0.5]{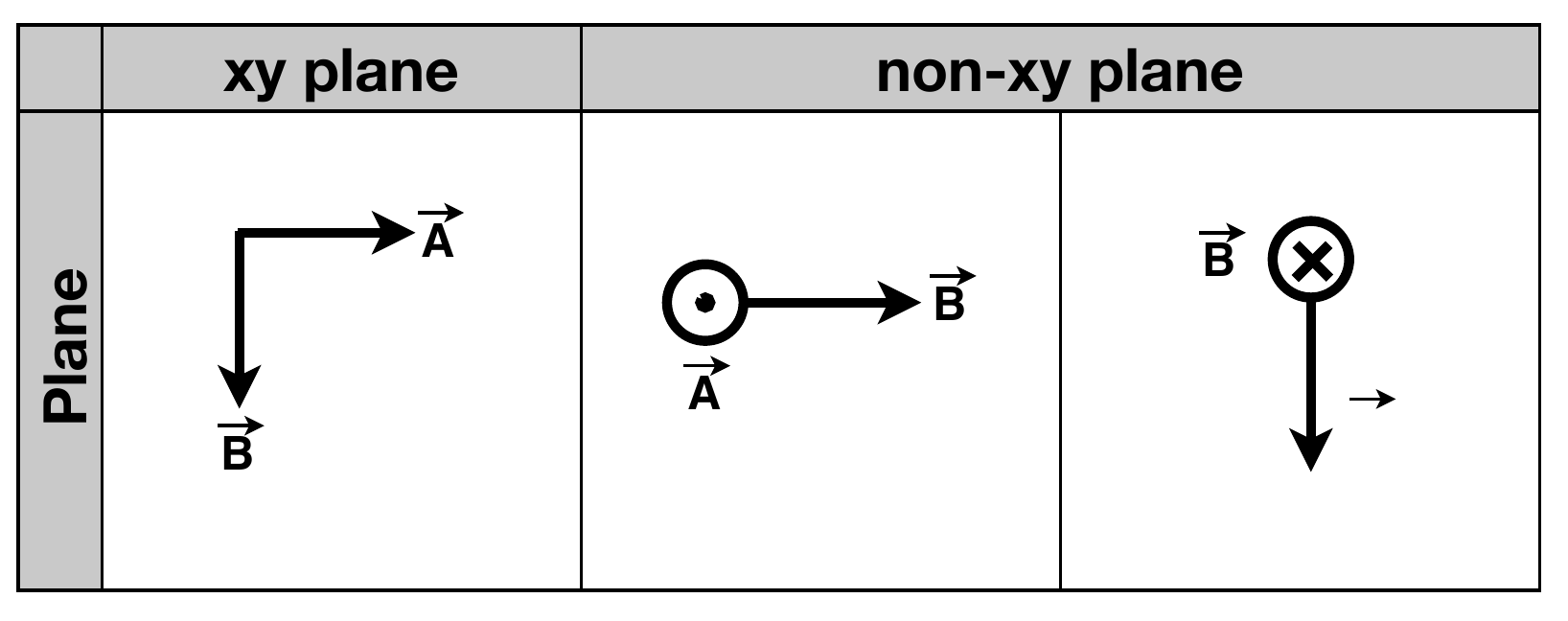}
\caption{Examples of physically awkward orientations in the $xy$ and non-$xy$ planes.}
\label{plane}
\end{center}
\end{figure}

Among the ten orientations identified as \emph{Easy} or \emph{Awkward}, there were physically easy orientations in the $xy$ and $yz$ planes and physically awkward orientations in the $xy$, $yz$, and $xz$ planes. Thus, combining the $yz$ and $xz$ orientations into ``non-$xy$'' orientations allows us to separately consider of the relative impact of the plane of the vectors and physical discomfort on performance. As an example, Fig.~\ref{plane} shows physically awkward orientations in the $xy$ and non-$xy$ planes.

\begin{figure}
\begin{center}
\includegraphics[scale=0.5]{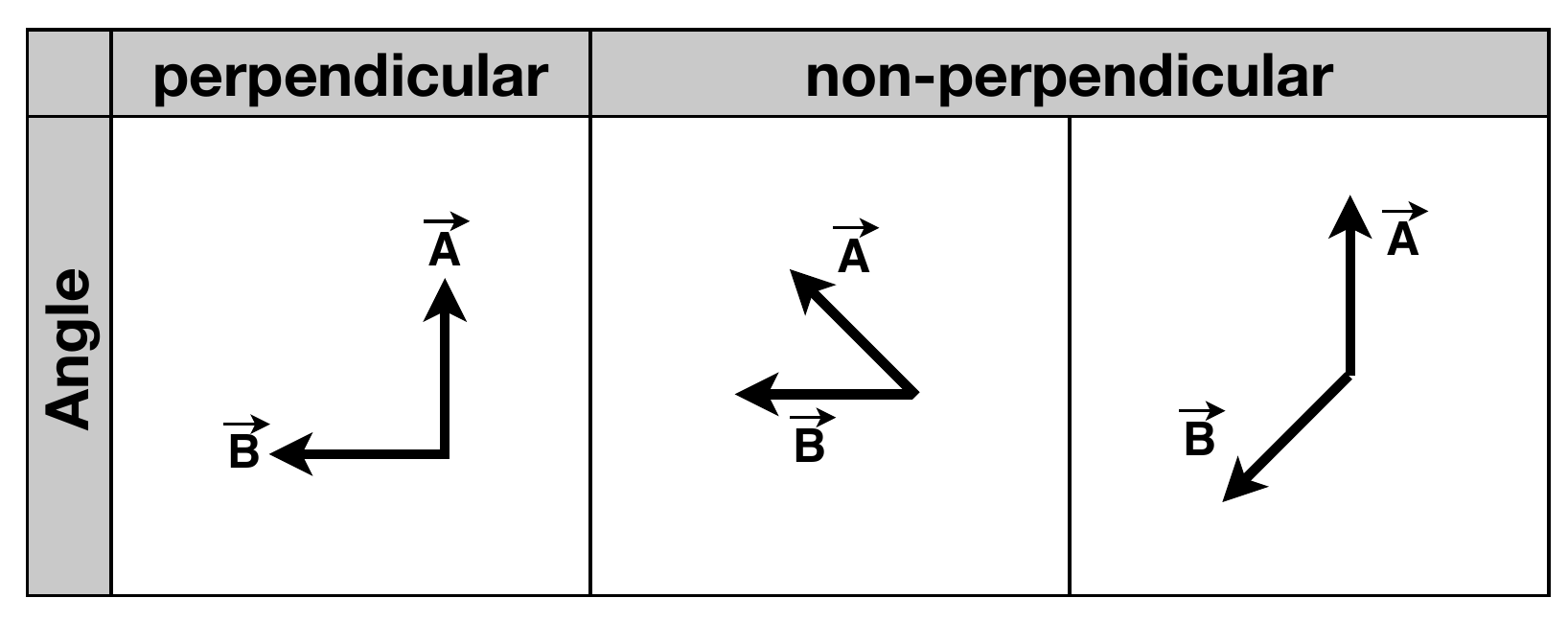}
\caption{Examples of one orientation, $(+\hat{y})\times(-\hat{x})=+\hat{z}$, with the vectors presented as perpendicular and non-perpendicular (with both acute and obtuse angles). }
\label{angle}
\end{center}
\end{figure}

In order to limit the interview to one hour, the angle between the vectors was only varied for the forward reasoning questions in a non-physics context. Also, to keep orientations similar to those that students might see in typical physics problems, only orientations in $xy$~plane were varied with respect to angle. There was a mix of acute and obtuse angles (combined into a \emph{non-perpendicular} category) that were compared to the majority of problems which had \emph{perpendicular} orientations. Figure~\ref{angle} shows the same basic orientation with perpendicular, acute, and obtuse angles.

\subsubsection{Physics features: `Charge', `Shape', and `Field'}
\begin{figure}
\begin{center}
\includegraphics[scale=0.5]{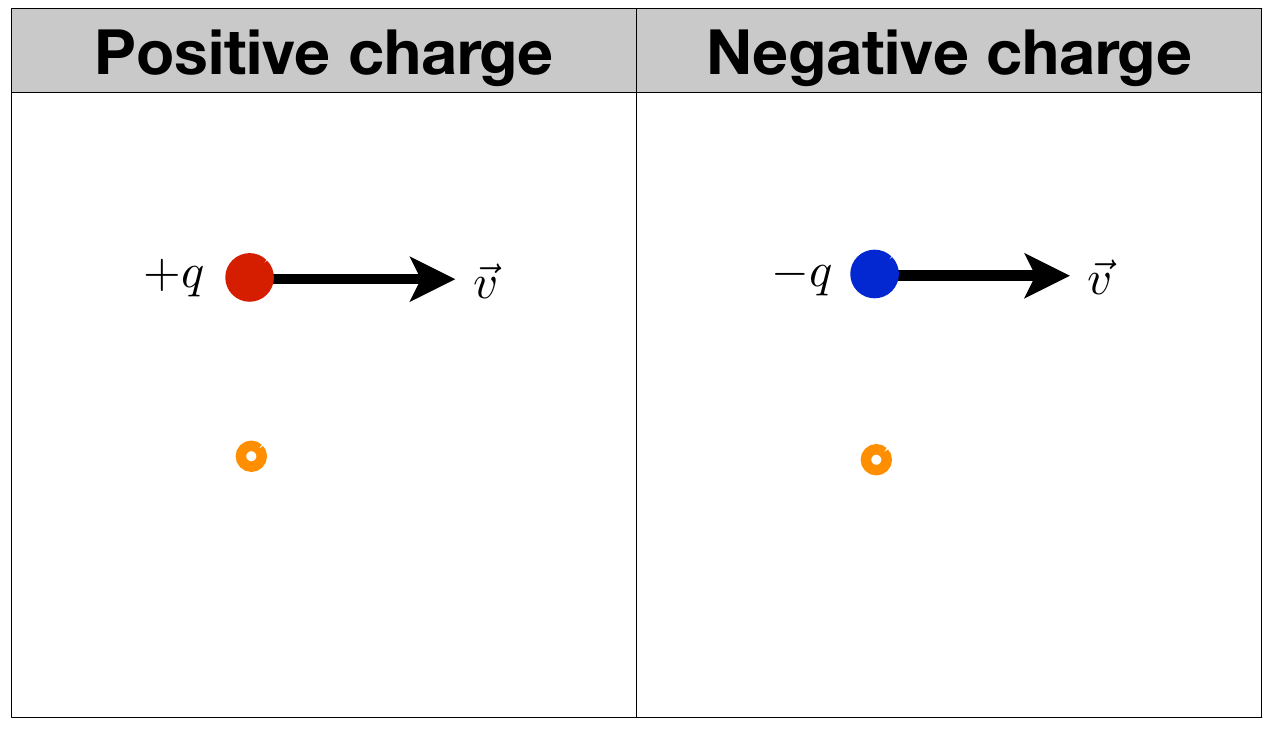}\\
(a)\\

\includegraphics[scale=0.5]{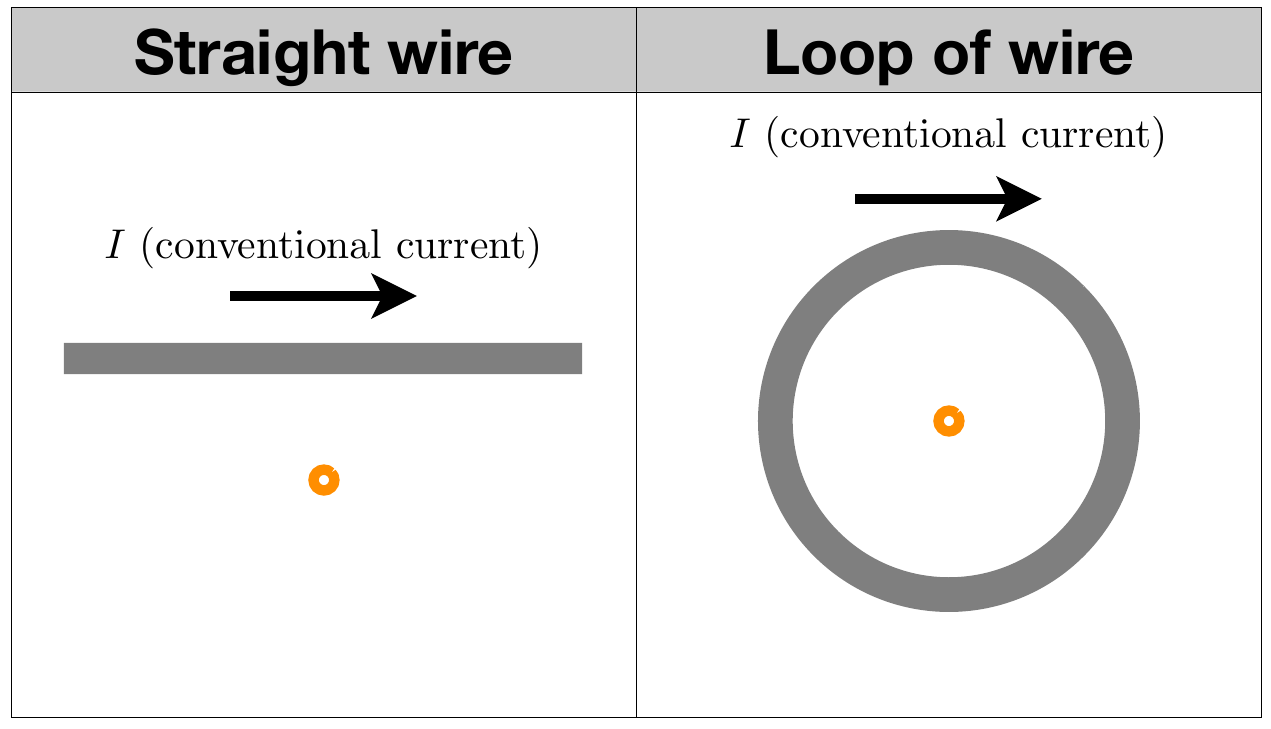}\\
(b)\\

\includegraphics[scale=0.5]{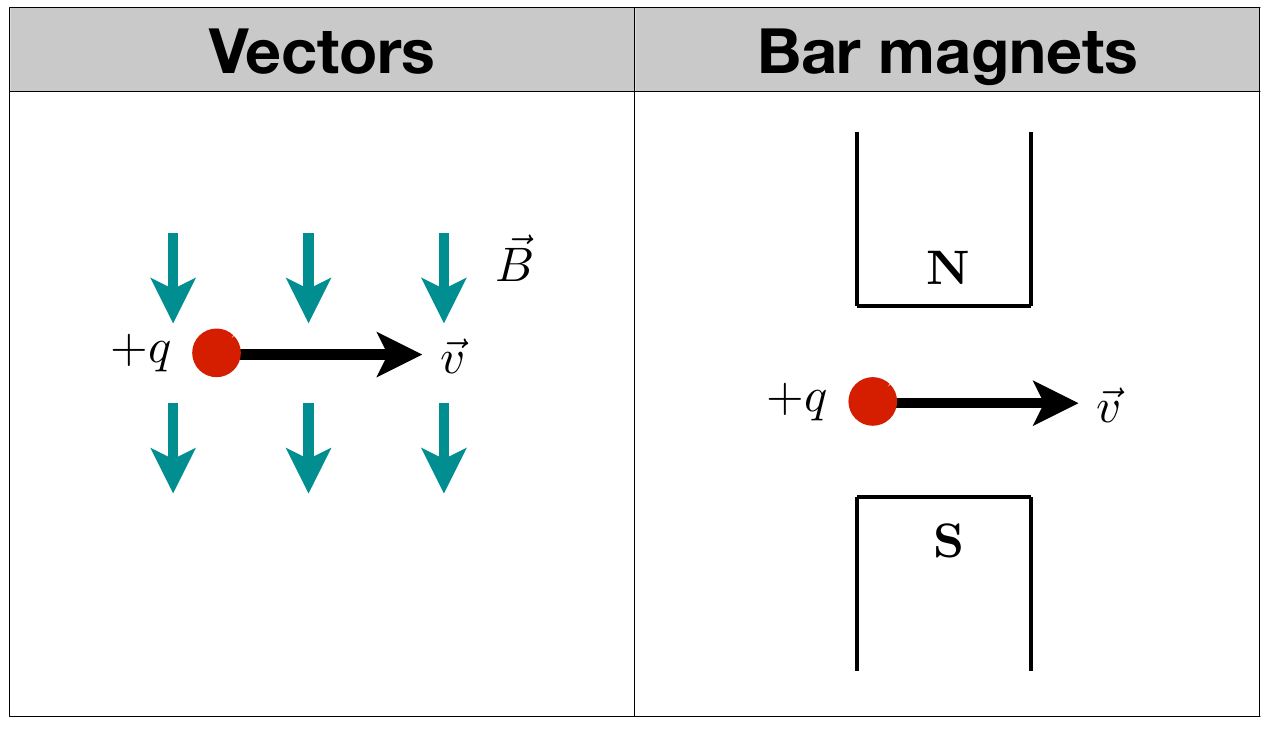}\\
(c)

\caption{Examples of the variation in physics features. The `Sign' of the charge, which was varied between the positive and negative charges in (a) and the current-carrying wires in (b). The `Shape' of the current-carrying wire in (b) were also varied between straight and circular. Finally, the representation of magnetic `Field' was varied between a vector/arrow representation and a bar magnet representation, shown in (c).}
\label{physics features}
\end{center}
\end{figure}

There were three main physics features: the sign of the charge (`Charge'), the shape of the current-carrying wire (`Shape'), and the representation of the magnetic field (`Field').
The variations in charge included a positive point charge (\emph{Positive}), a negative point charge (\emph{Negative}), and a current-carrying wire (\emph{Current}) with the direction of conventional current indicated. The shape of the current-carrying wire was only varied for the magnetic field problems and included straight wires (\emph{Straight}) and both square and circular current loops (\emph{Loop}). The representation of the magnetic field was only varied for the magnetic force problems and included either \emph{Vector} or \emph{Magnet}, analogous to the field lines and magnetic poles used in \citet{Scaife:2010}. Figure~\ref{physics features} shows examples of each of these features.

\subsection{Measures of performance}\label{performance}
In order to address the question of the how each feature discussed above will impact performance, one must first have a way to measure performance. For many problems, the most coarse measure of performance is whether the question was answered correctly. Yet, even correctness is not as straightforward as one might expect. For example, how does one distinguish between answering a question correctly on the first attempt and answering correctly on a final attempt after taking multiple different approaches, some of which may be inappropriate? In addition, correctness may not be the most illuminating measure of performance. The current study used four different measures of performance in order to provide a more holistic perspective: correctness of the final response, response time, methods used, and types of errors made. 

The first two of these measures --- the correctness of the final response and the time taken to give that response --- provide quantitative data about performance. Correctness (or accuracy) and response time (or latency) are often used in the spatial cognition literature as correlated measures of cognitive difficulty. For example, it is assumed that the process of aligning reference frames comes at a measurable cost:  the more mis-aligned the reference frames, the less accurate the response 
and the longer the response time will be \footnote{This assumption was validated in all cited studies, excepting a speed-accuracy tradeoff in \citet{Pani:1996}}. 

Choosing the final response allowed for the possibility of participants changing their response throughout the problem and made correctness a binary variable. Problems where a participant did not provide a final response ($N=33$) were not included in the analysis. 

Response time was measured in seconds using video time stamps.  The `Start' was defined as the moment the participant first saw the question. This was done to address the fact that some participants would read the entire problem statement before trying to solve it, while others would jump into the problem and refer back to the problem statement only as needed. This definition of start time likely inflates the response time for the physics questions compared to the non-physics questions due to the amount of additional reading. The operational definition for the `End' was the first instance of the participant's final response.

Analysis of this data involved regression analyses for correctness and response time to test the hypothesized impact of the problem features discussed in Table~\ref{hypotheses}. The details of these analyses and their results are discussed in Section \ref{quantitative}.

The other two measures---the methods participants used to solve the problem and the types of errors that participants made while solving the problem---were identified from the data using an iterative  coding process.
 This process and the codes that emerged are discussed in Section~\ref{qualitative} (with more detailed code descriptions in the appendix). These qualitative measures provides some of the nuance that is lost with the coarse-grained measures of correctness and response time.

\section{Quantitative analyses}\label{quantitative}

\subsection{Overview of quantitative analyses}\label{quantitative overview}

The impact of the problem features on the quantitative measures of performance---correctness and response time---was assessed using regression analyses. As defined in Section~\ref{performance}, correctness is binary and thus logistic regression was appropriate. In contrast, response time is a continuous variable with a Poisson-like distribution, so a standard least squares regression analysis was used with the logarithm of response time as the dependent variable (log-linear regression).

Since the focus of this study is the use of RHRs, problems where a participant did not use a RHR were eliminated from the analyses (these problems were identified using the qualitative coding discussed in Section~\ref{qualitative}). After these problems had been eliminated, $N=2119$ problems remained for use in the analyses.

Each regression included ten predictors: the first six problem features in Table~\ref{hypotheses}, the three control variables at the bottom of Table~\ref{hypotheses}, and the other quantitative measure of performance. As mentioned previously, it is a standard assumption in spatial cognition research that accuracy and latency are correlated measures of cognitive difficulty. To validate this assumption for the current study, correctness and response time were used as predictors for each other in the respective regression analyses. The results from the control variables will be presented as a part of the regression models for completeness, but the discussion here will focus only on the six problem features \footnote{A more detailed discussion of the control variables can be found in \citet{Kustusch:dissertation}}.

The remaining four problem features from Table~\ref{hypotheses} were not included in the regression analyses because they were not varied on all problems. The physics features (`Charge', `Field', and `Shape') could not be varied the same way across all contexts and in order to keep the interview to a reasonable length, `Angle' was only varied for non-physics working forward problems. To assess the impact of these four features, the author performed an analysis of variance (ANOVA) $F$ test for response time and a Likelihood Ratio $\chi^2$ test for correctness for the relevant subset of problems where each feature was varied.

It is important to note that the data are such that a strict interpretation of the results is impossible (\emph{e.g.}, not all features were varied on all problems) and that these regression analyses are designed to provide a first-order assessment of how individual problem features impact performance on problems requiring a RHR.  

\subsection{Results of quantitative analyses}\label{quantitative results}

\begin{table}
\caption{Logistic regression for correctness. The odds are for correct versus incorrect, where odds $>1$ indicate a positive impact on correctness and odds $<1$ indicate a negative impact on correctness. For continuous terms, the unit odds are per unit change in contrast and the range odds are per change in contrast over entire range. For categorical terms, odds are for Level 1/Level 2, i.e., for odds $>1$, the Level 1 category is more likely to be correct than the Level 2 category}
\label{correctness model}
\begin{ruledtabular}
\begin{tabular}{lr}
Model level statistics\\
\hline
\\
$R^2$ $(U)$					&0.1254\\
Observations ($N$)				&2119\\
$\chi^2$ statistic $p$ value			&$<0.0001$\\
\end{tabular}

\vspace{11pt}

Continuous terms\\

\begin{tabular}{lllr}
Term	&Unit Odds	&Range Odds	& $p$ value\\
\hline
\\
Log (T)			&0.53		&0.042		&$<0.0001$\\
Spatial			&1.0			&4.6			&$<0.0001$\\
Order			&1.0			&1.3			&\hspace{0.15in}$0.46$\\
\end{tabular}
\vspace{11pt}

Categorical terms\\

\begin{tabular}{lllll}
Term	&Level 1		&Level 2		&Odds	& $p$ value\\
\hline
\\
Experience	&1st group	&2nd group		&1.2		&0.12\\
Context		&Physics			&Non-physics		&0.96	&0.86\\
Type			&Forward			&Backward		&1.6		&0.0047\\
Comfort		&Easy			&Awkward		&1.5		&0.0070\\
Plane		&xy				&Non-xy			&1.5		&0.0046\\
Tails			&Tail				&Head			&1.0		&0.84\\
Separation	&Together			&Separate			&1.1		&0.72\\
 \end{tabular}
\end{ruledtabular}
\end{table}

\begin{table}
\caption{Least squares regression for Log (T). A positive estimate indicates that the category in square brackets contributed to a longer response time than the other category associated with that term. For example, the positive estimate for Type[Backward] implies that Backward problems took more time than Forward problems, whereas the negative estimate for Correctness[Correct] implies that Correct problems took less time than Incorrect problems. }
\label{time model}
\begin{ruledtabular}
\begin{tabular}{llr} 
Model level statistics\\
\hline
\\
$R^2$ (adjusted)		&0.7080&\\
Observations ($N$)		&2119&\\
$F$ statistic $p$ value	&0.0000&\\
\\
\hline
\hline
Term						&Estimate		&$p$ value\\
\hline
\\
Intercept					&+3.5		&0.0000\\
Correctness [Correct]			&$-$0.087	&$<0.0001$\\
Spatial					&$-$0.013	&$<0.0001$\\
Order					&$-$0.013	&$<0.0001$\\
Experience [1st group]	&+0.10		&$<0.0001$\\
Context [Magnetic field]		&+0.50		&$<0.0001$\\
Context [Magnetic force]		&+0.48		&$<0.0001$\\
Type [Backward]			&+0.18		&$<0.0001$\\
Comfort [Awkward]			&+0.032		&0.0060\\
Plane [Non-xy]				&+0.094		&$<0.0001$\\
Tails [Head]				&+0.012		&0.4246\\
Separation [Separate]		&+0.032		&0.0381\\
 \end{tabular}
\end{ruledtabular}
\end{table}

The logistic regression for correctness is presented in Table~\ref{correctness model}, including model level statistics and odds ratios. For logistic regression, $R^2$ or $U$ is the proportion of the uncertainty for the Whole Model Fit. For this model, there is only 12\% uncertainty attributed to the fit. 
Odds ratios are included since parameter estimates are not as easily interpreted for logistic regression. Odds greater than one indicate a positive impact on correctness and odds less than one indicate a negative impact on correctness. For continuous terms,  unit odds represent the impact on correctness per unit change in the term, while the range odds indicate the effect on correctness per change in the term over the entire range. For categorical terms, the odds specify the probability of getting a correct response with the first level divided by the probability of getting a correct response with the second level (i.e., for odds $>1$, the Level 1 category is more likely to be correct than the Level 2 category).

The odds for all contrasts were in a direction consistent with the hypotheses in Table~\ref{hypotheses}, with the exception of `Experience' and this may be due to interference effects between concepts of magnetic field and magnetic force (discussed in the qualitative analysis in Section~\ref{qualitative discussion}). However, the impact of most of these predictors on correctness was not as significant as their impact on response time (see Table~\ref{time model}). The problem features that appeared to have the largest impact on correctness were the type of reasoning required by the problem, the physical discomfort of the orientation and the plane of the given vectors.

The log-linear regression for response time is presented in Table~\ref{time model}, which includes model level statistics and parameter estimates. This model accounted for approximately 70\% of the variance in response time. Although it is difficult to strictly interpret significance due to the limitations of the data, the results show that all of the main effects impacted  response time in a direction consistent with the hypotheses in Table~\ref{hypotheses} and all were significant at the $p<0.01$ level except for Tails and Separation, the two features associated with parallel transport. 

As discussed in Section~\ref{quantitative overview}, there were several features which were not varied across all problems (`Angle', `Charge', `Shape', and `Field') and thus, not included in the regression analysis. To assess the impact of each of these features, the author performed an analysis of variance (ANOVA) $F$ test for response time and a Likelihood Ratio $\chi^2$ test for correctness for the relevant subset of problems where that feature was varied. The relevant subset for each of these features was:
\begin{itemize}
\item `Angle': all working forward, non-physics problems in the $xy$-plane ($N=723$);
\item `Charge': all magnetic field and magnetic force problems ($N=533$);
\item `Field': magnetic force problems where the magnetic field was represented ($N=105$); and
\item `Shape': magnetic field and magnetic force problems with a current-carrying wire ($N=160$).
\end{itemize}

 The $p$-values for each of these analyses are shown in Table~\ref{non-regression results}. As expected, for different shapes of current-carrying wires, there was no significant difference in correctness ($p=0.33$) or response time ($p=0.27$). For the other features, results were in a direction consistent with the hypotheses in Table~\ref{hypotheses}, but  differences in correctness ($\chi^2$) were significant at $p<0.05$ level whereas differences in response time ($F$) were not.

\begin{table}
\caption{Results for the analysis of variance (ANOVA) $F$ test for response time and Likelihood Ratio $\chi^2$ test for correctness for four features (Angle, Charge, Shape, and Field). The $N$ value reflects the number of relevant problems for which each feature was varied.}
\label{non-regression results}
\begin{ruledtabular}
\begin{tabular}{llll}
&&$F$ test& $\chi^2$ test\\
Term	&$N$	&$p$ value&$p$ value\\
\hline
\\
Angle&723&0.14&0.041\\
Charge&533&0.071&0.035\\
Field&105&0.34&0.0025\\
Shape&160&0.27&0.33
\end{tabular}
\end{ruledtabular}
\end{table}

\subsection{Discussion of quantitative analyses}\label{quantitative discussion}
The results from the quantitative analyses were generally consistent with expectations based on the literature and conventional wisdom. This section will briefly address the few exceptions as well as highlight points of interest that warrant particular notice in the context of the questions guiding this study.

\subsubsection{Kind of question and physics features}

The first thing of note is that the type of reasoning required (`Type') had one of the most significant impacts on both response time ($p<0.0001$) and correctness ($p=0.0047$). As expected, problems that require reasoning backward are significantly more difficult than those that require forward reasoning. Given its significance, the qualitative discussion (Section~\ref{qualitative discussion}) explores the interactions between this feature and the qualitative measures of performance.

In contrast with `Type', the physics context of the problems (or lack thereof) was the least significant of the predictors for correctness ($p=0.86$) and, as mentioned in Section~\ref{performance}, the results for response time are likely inflated due to the way response time was measured (including the time to read the problem). 

The results for `Context' are in contrast to \citet{VanDeventer:thesis}, who found that on cross product direction questions, the percentage of correct answers was significantly higher in a physics context (torque) than in a non-physics context. However, \citeauthor{VanDeventer:thesis} also found that the overall correctness on cross product direction questions was very low ($<30\%$). Thus, the lack of significance seen here may be due to an overall improvement in understanding about cross products from the first semester to the second semester of the introductory sequence. 

Alternatively, the lack of significance may be due to the difference in physics context (torque vs. magnetism). \citet{VanDeventer:thesis} found that in the torque context, the most common incorrect answer was that the direction was counterclockwise (consistent with the direction the pulley would move due to the torque applied). Since students have much more everyday experience with torque concepts than they do with magnetism concepts, it may be that the torque context cues inappropriate resources that the magnetism and non-physics contexts do not.

\subsubsection{Parallel Transport: Tails and Separation}
One of the more surprising results of this quantitative analysis is that the features associated with parallel transport appeared to have the least impact on both correctness and response time.
Despite literature showing that students struggle when vector operations require moving the vectors (\emph{e.g.}, \citet{Hawkins:2009}), it appears that parallel transport may not be as much of an issue for cross products as it is for vector addition and subtraction.

\subsubsection{Orientation}
As discussed in Section~\ref{spatial}, \citet{Scaife:2010} did not see any measurable effects on RHR use due to vector orientation, yet research in spatial cognition (e.g., \citet{Klatzky:2008}) strongly suggests that the orientation of the vectors should have an impact on performance. However, the spatial cognition literature also recognizes that orientation is not a single construct. By considering the physical discomfort, the plane of the vectors and the angle of the vectors as separate features, we see results that are consistent with the spatial cognition literature.

The impact of both physical awkwardness (`Comfort') and the plane of the given vectors (`Plane') was significant at the $p<0.01$ level for both correctness and response time, indicating that these features separately contribute to the difficulty of using a RHR. However, it is important to note that this study used the most physically easy and the most physically awkward orientations. Therefore, it is possible that the effect from physical discomfort is inflated. Also, as will be discussed in 
the qualitative analysis (Section~\ref{qualitative discussion}), 
difficulty with vectors that are not in the xy-plane may also be due to 
issues other than mental rotation.

The angle between the vectors (`Angle') also had a significant impact on correctness ($p=0.041$), but not on response time ($p=0.14$). The lack of a significance on response time is likely due to the fact that the angle was varied only for non-physics working-forward problems in the xy-plane. Based on correctness and response time, this subset of problems is one of the of the easiest (the overall level of correctness for this subset was above $90\%$). Since there is a significant difference in correctness for even these easy problems, it is possible that on more difficult problems the angle would have an even stronger effect. Additionally, angle plays a key role in geometric methods for finding the magnitude of a cross product, which is not discussed here. Thus, the angle of the vectors should be considered in any future study examining student performance on cross product problems.

\subsubsection{Physics features}\label{quantitative physics}

The results from the physics features (`Charge' and `Field') demonstrate that within a physics context, the physical features of the situation do add to the challenge of appropriately using a RHR. Both the sign of the charge and the representation of the field had a significant impact on correctness ($p<0.05$), although not on response time. This difference in significance could indicate a speed-accuracy tradeoff. For example, if participants do not acknowledge the need to treat a negative charge differently than a positive charge, they will answer questions with negative charges as quickly as those with positive charges, but will answer them incorrectly.

\subsection{Summary of quantitative analyses}
The primary goal of the quantitative analysis was to provide a first-order look at the relationship between the features identified in the literature and the use of RHRs on cross product direction questions, as measured by correctness and response time. The results of these analyses show that each feature does impact performance in a direction consistent with the hypotheses (Table~\ref{hypotheses}) and while it is difficult to strictly interpret significance, some features have a greater effect than others.

The type of reasoning required (`Type'), all three aspects of orientation (`Comfort', `Plane', and `Angle'), and certain physics features (`Charge' and `Field'), all have a significant impact on student performance. Yet, it appears that features involving parallel transport (`Tails' and `Separation') may not have the same kind of impact on cross product problems that they do on vector addition and subtraction problems. Finally, understanding the role of `Context' will require more study to differentiate whether the differences between results here and previous research are due to an increased familiarity with cross products or with the specifics of the physics context.

\section{Qualitative Analyses}\label{qualitative}

\subsection{Overview of qualitative analyses}\label{qualitative overview}

As discussed in Section~\ref{performance}, correctness and response time are fairly coarse measures of performance. Additional insight can be gained by a more qualitative look at student use of RHRs. An iterative coding process was used to identify the methods that participants used to solve the problems and the types of errors that they made while doing so.

An initial set of codes were derived using content logs of interviews, as well as a pilot study conducted the previous semester. These codes were anchored in published research, but were then revised by a constant comparison with the data to determine if the codes fit the data. Once this revision process reached a point where it appeared that the codes were an accurate reflection of the data, a second independent coder was brought in to determine the reliability of the code definitions. This process led to further revisions to clarify definitions and further rounds of reliability testing. Each round of inter-rater reliability used 3-8\% of the more than 2400 questions. To compensate for the smaller data set and to provide for more robust agreement, the problems used for interrater reliability were chosen to reflect the most difficult questions and the widest variety of codes. Since neither methods nor errors are mutually exclusive, each problem was given a code for every method and for each family of errors.
  Cohen's kappa was calculated separately for each code family (ranging from 0.7-1.0). \citet{Kustusch:dissertation} provides a more detailed description of the development and validation of these coding schemes.
  
Once all problems were coded for both methods and errors, the relationships between these methods and errors were explored, as well as how these methods and errors interacted with the various problem features of the tasks. Given the number of codes and problems, the results of the quantitative analysis were used to focus this exploration on those problem features that had the most impact on correctness and response time. The following sections will present the most interesting and relevant of these codes and interactions, but a more complete analysis can be found in \citet{Kustusch:dissertation}.

\subsection{Results of qualitative analyses}\label{qualitative results}

 \begin{figure}
\begin{center}
\includegraphics[width=3.5in]{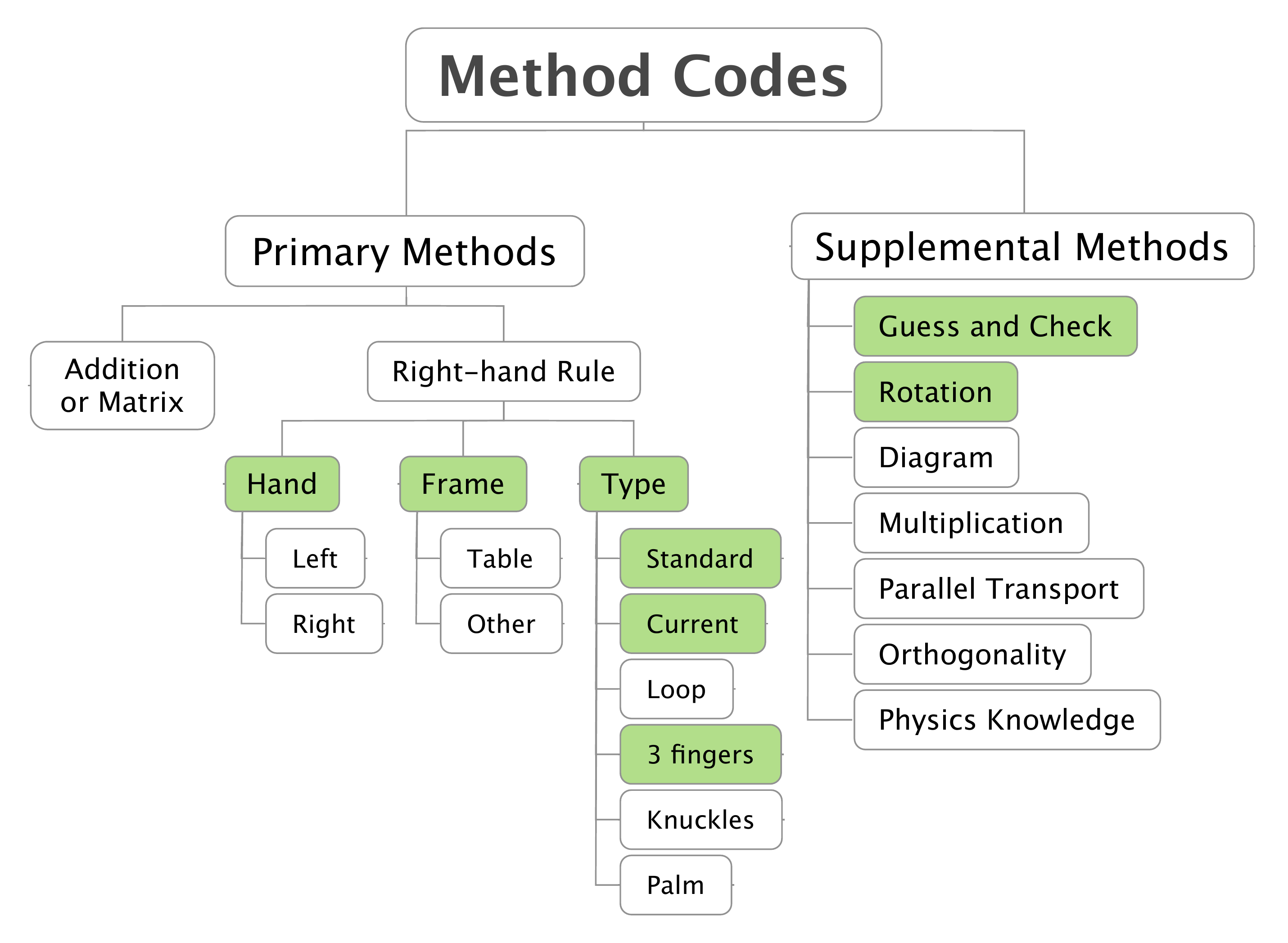}
\caption{Graphical depiction of method codes and sub-codes. Those in green are discussed in the text and in Table~\ref{method code definitions brief}.}
\label{method codes map}
\end{center}
\end{figure}

\begin{table}
\caption{Brief definitions for  methods discussed in the text. Descriptions of all methods can be found in Table~\ref{method code definitions full}.}
\label{method code definitions brief}
\begin{ruledtabular}
\begin{tabular}{p{0.8in}p{2.45in}}
Code Name& Code Definition\\
\hline
\\
\emph{RHR:Hand} & Specifies whether participant uses the `Right' or `Left' hand\\
\\
\emph{RHR:Frame}
& Specifies whether the right-hand rule is performed in the plane of the table (`Table'), where $+ \hat{x}$ is to the right, $+ \hat{y}$ is away from the participant, and $+ \hat{z}$ vertically upward, or in a different (`Other') frame, e.g.,  $+ \hat{z}$ is toward participant instead of vertically upward.\\
\\
\emph{RHR:Type:} `Standard'&Fingers point in the direction of the first vector ($\vec{A}$), curl in the direction of the second vector ($\vec{B}$) and the thumb points in the direction of $\vec{A}\times \vec{B}$.\\	
\\
\emph{RHR:Type:}  `Current'&Thumb points in the direction of conventional current and the fingers curl such that at a given observation location, the fingers point in the direction of the magnetic field at that location.\\
\\
\emph{RHR:Type:} `3--Finger'& The index finger points in the direction of the first vector ($\vec{A}$), the middle finger points in the direction of the second vector ($\vec{B}$) and the thumb points in the direction of $\vec{A}\times \vec{B}$.\\	
\\

\emph{Guess-and-Check} & Participant explicitly ``guesses'' an answer or makes an assumption (e.g.,  a particle has positive charge) and uses another method to determine whether the assumption is consistent with the given information.\\
\\
\emph{Rotation} & Given frame of reference is rotated by more than $\approx15^\circ$ using the paper or a diagram.\\
\\
\end{tabular}
\end{ruledtabular}
\end{table}

The method codes and sub-codes are graphically depicted in Fig.~\ref{method codes map}. Those in green will be discussed more fully in the text and brief definitions of these are provided in Table~\ref{method code definitions brief}. Descriptions of all methods can be found in the appendix (Table~\ref{method code definitions full}).\

There were indications of two distinct families of methods: primary and supplemental. Primary methods can, on their own,  yield a singular, non-zero answer and the most common were the various RHRs. There was the occasional use of matrices, as well as vector addition and subtraction methods, but neither of these will be addressed in this paper. Supplemental methods are those that when used on their own cannot yield a singular, non-zero answer. They are the tools, techniques, and strategies that the participants usually used in conjunction with a primary method. For example, using the orthogonality of the cross product to restrict the solution to one axis and then using a RHR to determine the direction along that axis.

\citet{Greenslade:1980} points out that there are numerous physical mnemonics for the cross product that can vary widely. For the purpose of this study, all such mnemonics have been grouped as \emph{Right-hand Rules} (or \emph{RHR})  and identified with three sub-codes: the \emph{Hand} that was used (`Left' or `Right'); the \emph{Frame} of reference with which the hand was aligned (`Table' or `Other'), and the general \emph{Type}. For example, the RHR described by \citet{Klatzky:2008} is called the `Standard' right-hand rule. The two others that will be discussed in this paper are the `Current' rule, where the thumb points in the direction of the current and the fingers curl to the observation location, and the `3-Fingers' rule, where the index finger points along $\vec{A}$, the middle finger along $\vec{B}$ and the thumb along~$\vec{C}$ (see Table~\ref{method code definitions full} for descriptions of all types). If any of these sub-codes changed during the problem, it was considered a different use of a RHR. Thus, RHR was the only method code used more than once per problem.

The two supplemental methods that will be discussed in the following section are \emph{Guess-and-Check} and \emph{Rotation}. As the name indicates,  \emph{Guess-and-Check} involved making a guess or assumption about the answer and using another method (usually a RHR) to check if the guess was consistent with the given information. \emph{Rotation} involved rotating the given frame of reference, usually by physically rotating the paper, but also sometimes by redrawing the given diagram rotated along some axis.

\begin{turnpage}
 \begin{figure*}
\begin{center}
\includegraphics[scale=0.45]{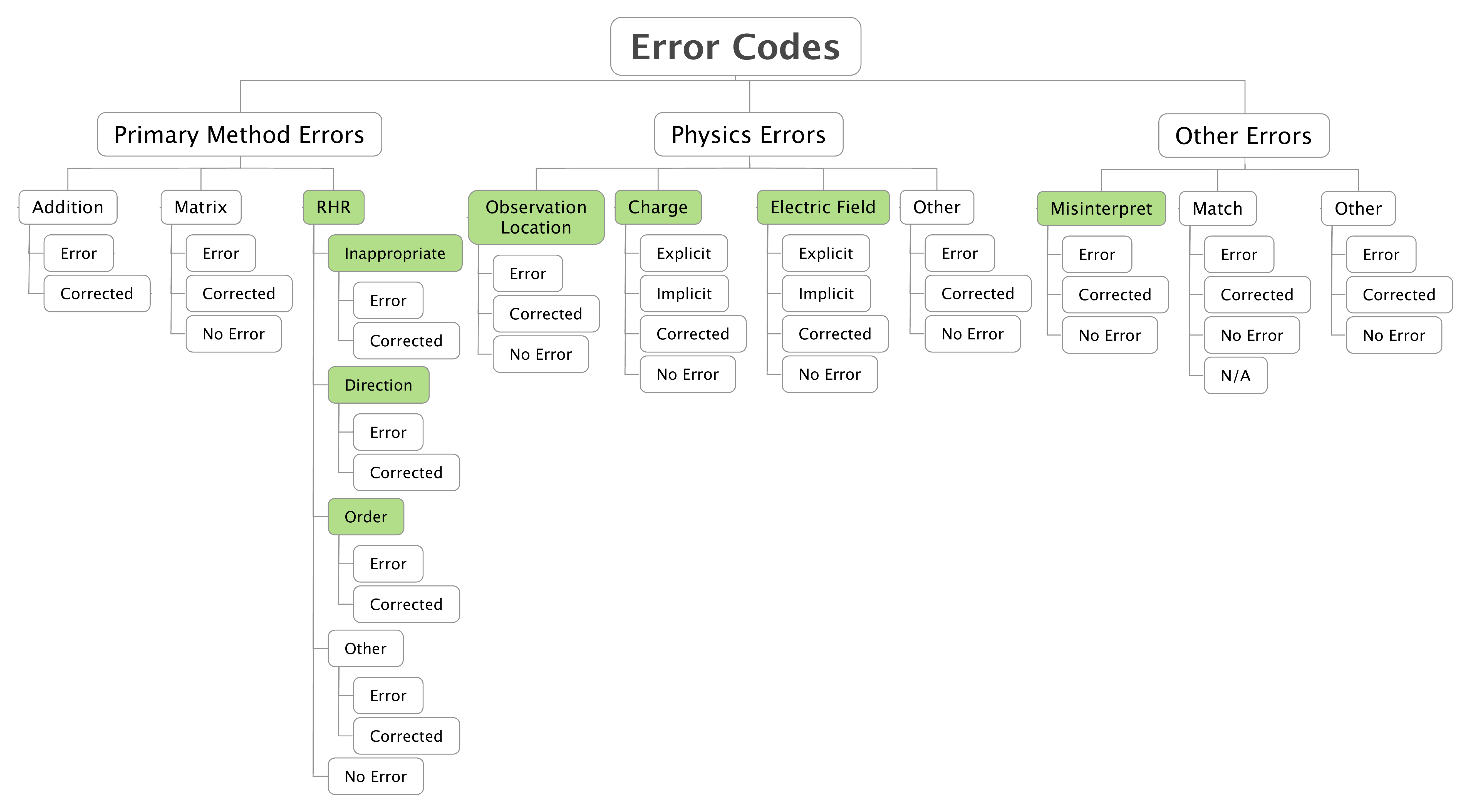}
\caption{Graphical depiction of error codes and sub-codes. Codes in green are discussed in the text and described in Table~\ref{error code definitions brief}}
\label{error codes map}
\end{center}
\end{figure*}
\end{turnpage}

\begin{table}
\caption{Brief definitions of the error types discussed in the text. Descriptions of all error types can be found in Table~\ref{error code definitions full}. The definitions here correspond to `Error' sub-codes, where sub-codes of `Corrected' or `No Error' are also possibilities. For \emph{Charge} and \emph{Electric Field}, explicit and implicit errors were categorized separately.}
\label{error code definitions brief}
\begin{ruledtabular}
\begin{tabular}{p{0.8in}p{2.55in}}
Code Name& Code Definition\\
\hline
\\
\emph{RHR: \hspace{0.3in}Inappropriate}	& Participant attempts to make use of a RHR that is inappropriate for the given situation. \\
\\
\emph{RHR: \hspace{0.3in}Direction}	 &Participant inappropriately reverses the direction of one of the given vectors. \\
\\
\emph{RHR: \hspace{0.3in}Order} & Participant performs the RHR in such a way that the order of the vectors is reversed and they are performing $\vec{B} \times \vec{A}$ instead of $\vec{A} \times \vec{B}$. 	\\
\\
\emph{Observation Location}& Participant does not appropriately account for the observation location. \\
\\
\emph{Charge} &Participant does not appropriately account for the sign of the charge. \\
\\
\emph{Electric Field}& Participant explicitly reasons from or provides an answer consistent with, an invalid or inappropriate analogy to electric field. \\
\\
\emph{Misinterpret}& Participant misinterprets the diagram or misreads the problems statement in a way that contributes to the final response. \\
\end{tabular}
\end{ruledtabular}
\end{table}

The full set of error types are shown in  Fig.~\ref{error codes map}. The codes in green are discussed in this paper and described in Table~\ref{error code definitions brief}. Brief descriptions for all codes can be found in the appendix (Table~\ref{error code definitions full}). Errors were grouped into three larger categories: errors associated with a primary method (only RHR errors will be discussed here), physics errors, and all other types of errors. Since the different types of errors are not mutually exclusive, every problem received an code for each type of error and each unique use of a RHR received a corresponding error code. 

When an error type contributed to the final response, it was coded as `Error'. However, there were times when an error was made, but it did not contribute to the final response. This typically occurred because the participant recognized the error and corrected it or abandoned the approach that led to that error --- these situations were coded as `Corrected'. A code of `No Error' indicated that there was no evidence for that type of error for that problem.  Thus, a problem that received `No Error' codes for all types was answered correctly, although the converse was not necessarily true due to corrected errors and multiple errors that canceled each other out. 

For some types of errors, there were examples with implicit evidence (\emph{e.g.,} not accounting for the sign of a negative charge) and examples with explicit evidence (\emph{e.g.,} explicitly double counting the negative sign for problems involving an electron). For \emph{Charge} and \emph{Electric Field} errors, implicit and explicit error were categorized separately, but for \emph{Observation Location} errors, explicit and implicit errors were grouped together.

Based on the coding described here, patterns emerged  that were associated with certain errors, methods and/or problem features. The most relevant of these patterns will be presented and discussed in the following section.

\subsection{Discussion of qualitative analyses}\label{qualitative discussion}
This section discusses five aspects of the qualitative coding and analyses that build on the quantitative analyses in Section~\ref{quantitative} by providing:
\begin{enumerate}
\item a more nuanced perspective on student difficulties with the commutativity of the cross product,  
\item an alternate explanation for student difficulties with non-$xy$ plane orientations,
\item a look at a supplemental method (\emph{Rotation}) that appears to mitigate difficulties with orientation, 
\item the verification of several previously identified difficulties with physics content, and 
\item the identification of physics difficulties that have not previously been discussed in the literature. 
\end{enumerate}
These outcomes demonstrate the complexity of how contextual and representational features impact student use of right-hand rules.

\subsubsection{Adding nuance to the issue of commutativity\label{RHR order}}

The largest number of uncorrected errors associated with a RHR were \emph{Order} errors. This is a type of error where the participant either implicitly or explicitly reverses the order of the vectors (i.e., $\vec{C}=\vec{B}\times \vec{A}$ instead of $\vec{C}=\vec{A}\times \vec{B}$). The existence and prevalence of this kind of error gives credence to the claim by \citet{Scaife:2010} that one of the causes of ``sign errors'' is that students do not recognize the non-commutativity of the cross product. However, in their study, sign errors committed by students who treated the cross product as commutative appeared to non-systematic. By looking at the \emph{Order} errors in the current study more closely, we can illuminate the complexity of this issue of commutativity.

As mentioned in Section~\ref{study context} the interviewer  asked each participant  at the end of the interview how $\vec{B}\times \vec{A}$ compared to $\vec{A}\times \vec{B}$. Many participants stated that these two quantities would have the same magnitude, but point in opposite directions. Yet even these students committed \emph{Order} errors. So, while they may acknowledge the non-commutativity of the cross product when asked directly, they  still sometimes operationally treat the cross product as commutative. Examining the interaction between \emph{Order} errors and the various problem features reveals one possible reason for this disconnect.

\begin{figure}
\includegraphics[width=3.2in]{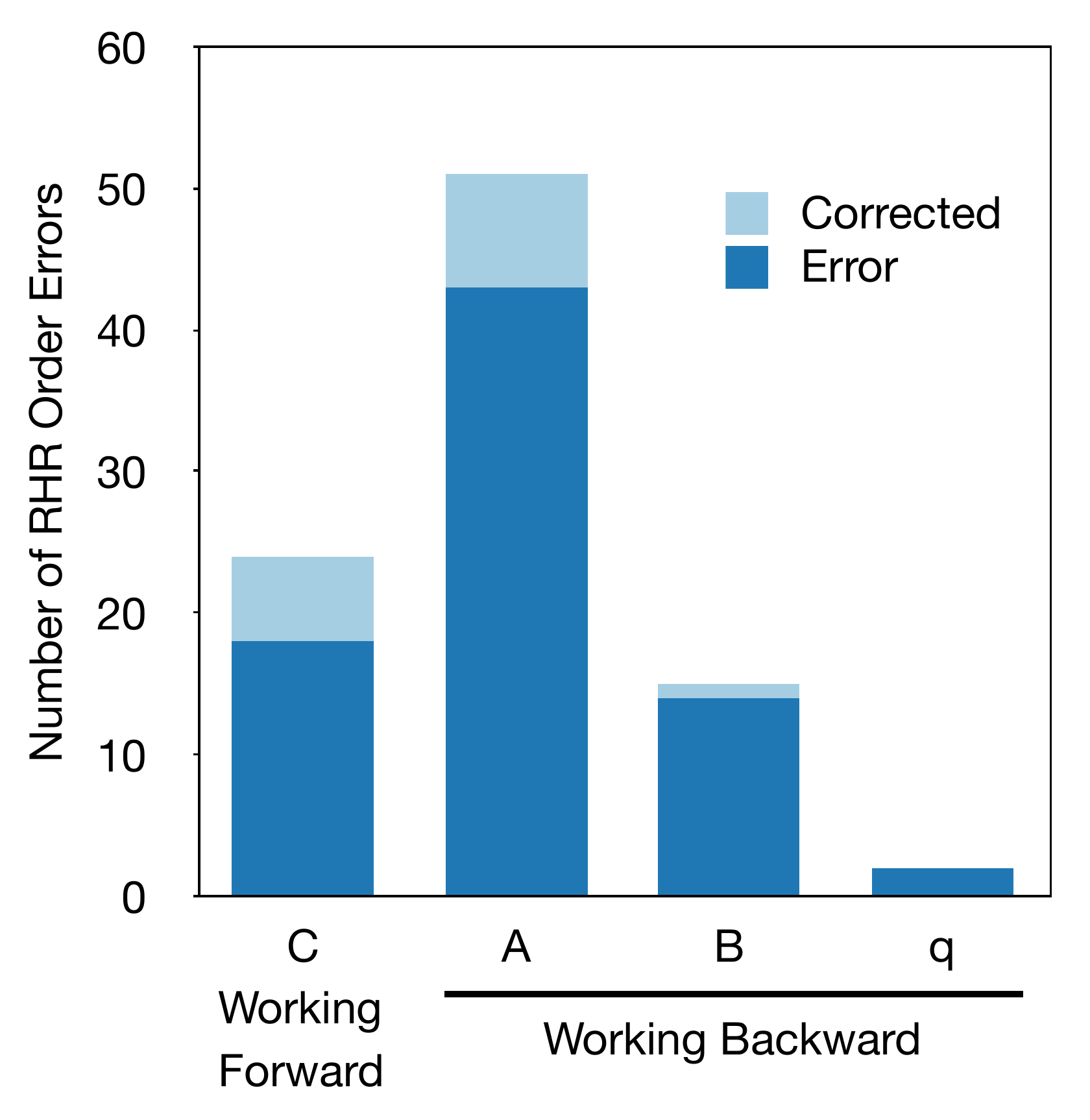}
\caption{The number of \emph{RHR: Order} errors by the type of reasoning required. \emph{Backward} problems are separated by the type of answer required, i.e., $A$ means that the first vector in the cross product was required, $B$ means the second vector was required, and $q$ means that all vectors were provided and the sign of the charge was required.}
\label{order errors by answer}
\end{figure}
\begin{figure}
\begin{minipage}{3.2in}
\smallskip
\begin{description}
\item[Problem] The magnetic field due to the negative charge is in the $-z$ direction (into the page) at the observation location shown. What is a possible direction for the velocity of the charge?
\end{description}
\bigskip
\includegraphics[scale=0.75]{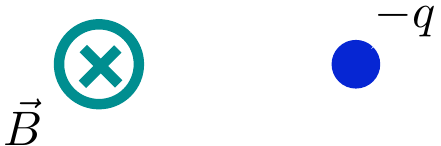}
\bigskip
\end{minipage}
\caption{A magnetic field problem that requires reasoning backward to find the first vector in the cross product.}
\label{order error}
\end{figure}

The quantitative results in Section \ref{quantitative} identified the type of reasoning as a feature with a notable impact on correctness and response time. Therefore, it should not be surprising that more than $70\%$ of the \emph{Order} errors were committed on Backward problems (see Fig.~\ref{order errors by answer}). What was slightly more surprising was that more than half of all \emph{Order} errors were on problems where the given vectors were $\vec{B}$ and $\vec{C}$ and the participant was asked to find $\vec{A}$, such as the one shown in Fig.~\ref{order error}. Even students who performed well on all problems ($>90\%$) and who explicitly identified the non-commutativity of the cross product committed this type of error (\emph{Order}) on this kind of working backward problem (finding $\vec{A}$).

Barry was one such participant who generally performed well. On the problem in Fig.~\ref{order error}, Barry made an \emph{Order} error, but then corrected himself. When asked what made the problem difficult, he stated,
\begin{quote}
``Gosh,
 it's hard to start without the initial direction, 
because what I see on the page is two points and so I'm thinking this [points fingers to the left] automatically, 
but really, that's the second step in finding a direction.''
\end{quote}
This `automatic' response that Barry identifies is consistent with the idea that problems which require working backward have a higher cognitive load than those that require working forward. It suggests that the \emph{Order} errors committed by those who do understand the non-commutativity of the cross product may be at least partially the result of not appropriately compensating for the backward nature of the problem.

It is interesting that when \emph{Guess-and-Check} was used as a supplemental method on working backward problems, the percentage of \emph{Order} errors was only $8\%$ compared to $15\%$ when \emph{Guess-and-Check} was not used. Additionally, \emph{Guess-and-Check} was used almost exclusively on Backward problems, suggesting that participants varied their methods in part to alleviate the challenges of this particular problem feature.

We can also look at individuals, such as Andrea, who explicitly stated that $\vec{B}\times \vec{A}=\vec{A}\times \vec{B}$ to better explore the non-systematic nature of these errors.
Examining the features of the problems where Andrea made \emph{Order} errors and those where she did not, it became clear that while her belief in the commutativity of the cross product contributed to these errors, the errors were by no means arbitrary. Every problem where she committed an \emph{Order} error was one where the orientation was physically awkward. Andrea consistently used the `3-Finger' RHR and on physically awkward problems, she would often reverse which finger represented which vector, essentially reversing the order of the cross product. For her, the belief that order did not matter did not create an arbitrary error, but one that was a way of dealing with the physical awkwardness of some of the orientations. This interaction between certain problem features and \emph{Order} errors for those that do not understand non-commutativity could partially account for the apparent non-systematic sign errors observed by \citet{Scaife:2010}.

The patterns discussed here demonstrate that while the connection between \emph{Order} errors and commutativity is real, this interaction is often mediated by problem features, such as the type of reasoning required and the physical awkwardness of using a RHR, and by other methods, such as \emph{Guess-and-Check}.

\subsubsection{Alternative explanation for non-$xy$ orientations \label{plane errors}}
As mentioned in Section~\ref{quantitative}, the results of the quantitative analysis show that when the orientation of the vectors is treated as more than a single construct, each aspect of orientation does impact student performance as we would expect based on the spatial cognition perspective. Based on literature on mental rotations, we expect that participants would perform better on problems in the $xy$ plane (where the axis of rotation is vertical) than on those in a non-$xy$ plane. While this trend is seen in the quantitative analysis, the qualitative analysis revealed an additional issue related to the plane of the vectors that is not directly related to mental rotations. 

There were two types of errors that were more common on $non-xy$ plane problems (\emph{i.e.}, one of the given vectors was into or out of the page in the $\pm z$ direction) than on $xy$ plane problems. One of these errors was the \emph{RHR: Direction} error, where the participant implicitly reversed the direction of one of the vectors while attempting a RHR. Due to the implicit nature of this error, it may be the participants are unaware that they are actually doing it. The other error was a \emph{Misinterpret} error, where the participant explicitly states a wrong direction, but performs a RHR appropriately for the stated direction. 

\begin{figure}
\begin{minipage}{3.2in}
\smallskip
\begin{description}
\item[Problem] What is the direction of $\vec{A}\times\vec{B}$?
\end{description}
\bigskip
\includegraphics[scale=0.7]{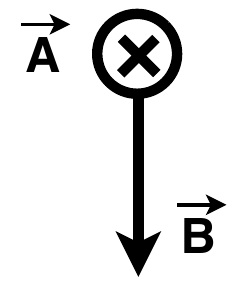}
\end{minipage}
\caption{An example problem where the given vectors are in the non-$xy$ plane.}
\label{misinterpret}
\end{figure}

Consider Fig.~\ref{misinterpret} to understand the difference between these two error types. If one were to use the `Standard' RHR on this problem, the fingers should point into the page and curl down. If the participant said that $\vec{A}$ was into the page, but pointed the fingers out of the page (reversing the direction of $\vec{A}$), this would be considered a \emph{Direction} error. If, however, the participant explicitly stated that $\vec{A}$ was out of the page and pointed the fingers out of the page, it would be coded as a \emph{Misinterpret} error.

The existence of \emph{Misinterpret} errors and their prevalence on problems where one of the vectors is along the $z$ axis suggests an additional reason for poorer performance on non-$xy$ plane problems. While some of the errors on these problems may be related to difficulties with mental rotations, there is likely also an issue related to understanding the $\bigotimes$ and $\bigodot$ symbols for into and out of the page, which are conventionally used to show a three-dimensional situation in a two-dimensional plane. With the current data, it is not possible to disentangle which is the more prominent source of difficulty, but it raises an issue that should be addressed in the future research using these symbols.

\subsubsection{Mitigating difficulties with orientation by rotation\label{rotation discussion}}
\begin{figure}[h]
\begin{center}
\includegraphics[width=3.2in]{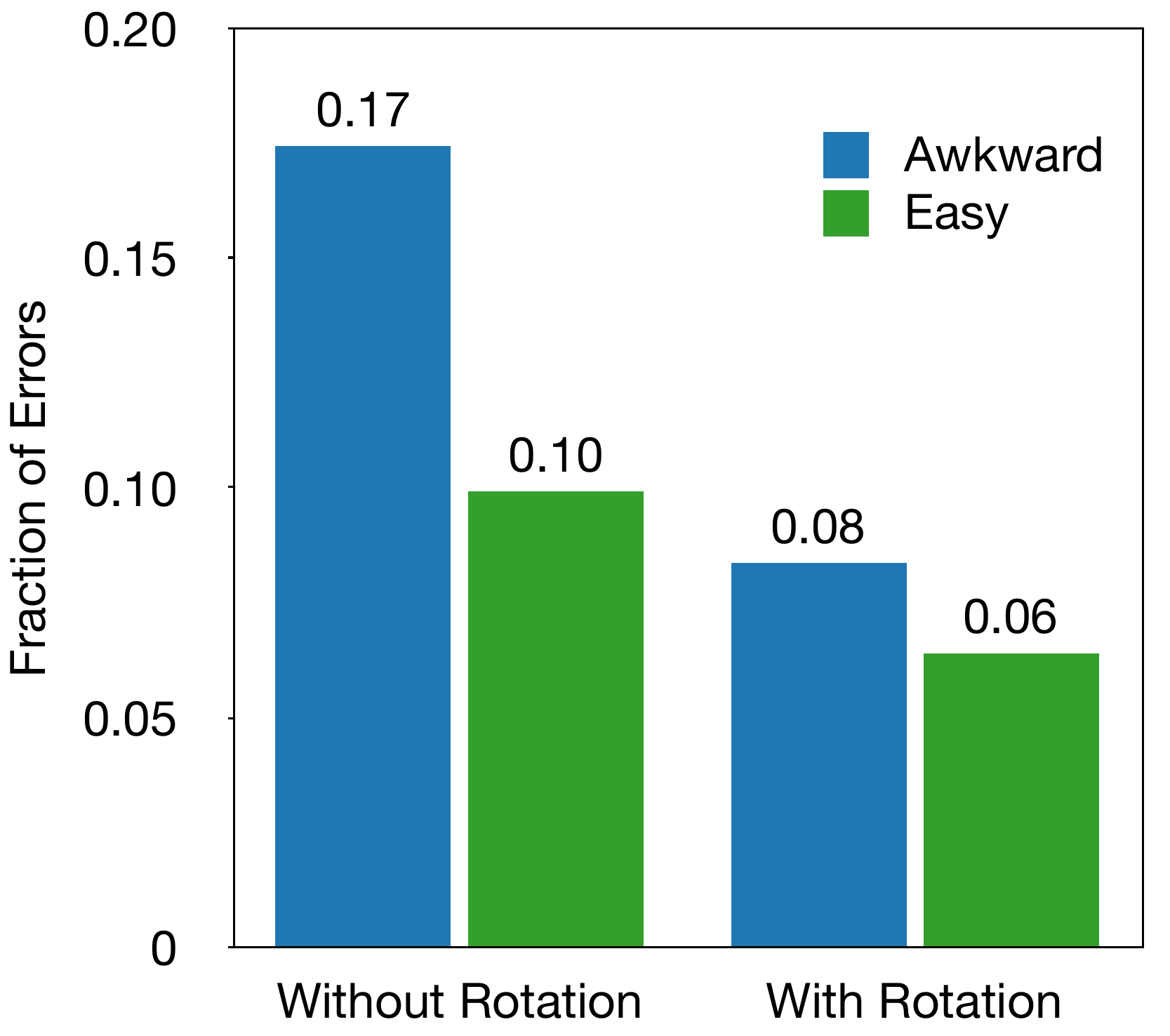}\\
(a)\\

\includegraphics[width=3.2in]{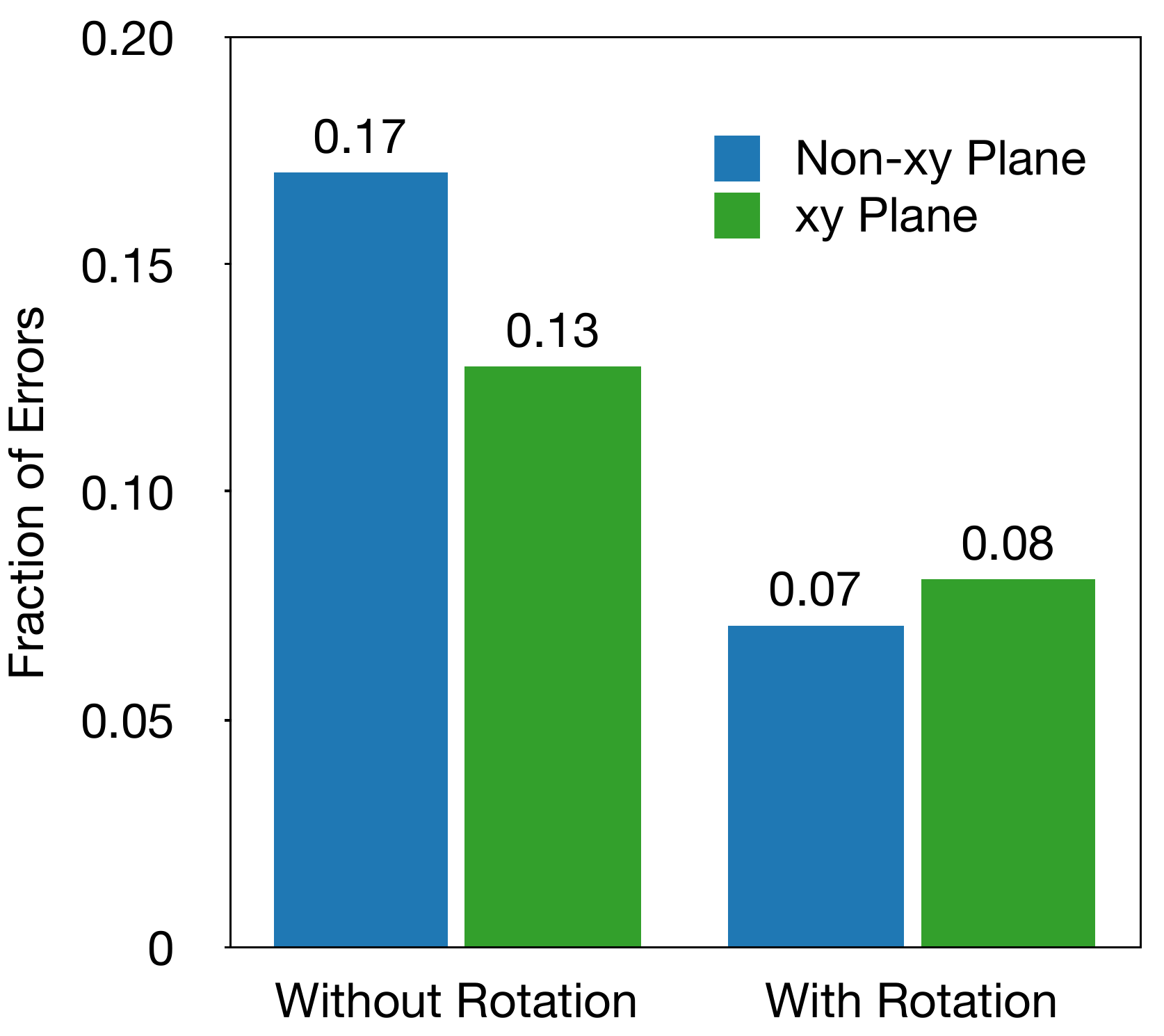}\\
(b)
\caption{Interaction of rotation and orientation: (a) fraction of RHR errors on physically easy and awkward problems with and without rotation; (b) fraction of RHR errors on problems in the $xy$ and non-$xy$ planes with and without rotation.}
\label{rotation}
\end{center}
\end{figure}

As mentioned above (Section~\ref{RHR order}), when students used the \emph{Guess-and-Check} method on Backward problems, the percentage of \emph{Order} errors was smaller than when they didn't use \emph{Guess-and-Check}. There was a similar effect from using \emph{Rotation} for two of the orientation features: `Comfort' and `Plane'. As shown in Fig.~\ref{rotation}, when \emph{Rotation} was used, the fraction of RHR errors is smaller overall compared to when Rotation was not used. In addition, the difference between \emph{Easy} and \emph{Awkward} problems is reduced (Fig.~\ref{rotation}.a), as is the difference between $xy$ plane and non-$xy$ plane problems (Fig.~\ref{rotation}.b).

Some instructors already encourage their students to physically rotate the paper if necessary to allow them to more easily perform a RHR. The findings presented here support the claim that this approach does indeed help to mitigate difficulties associated with orientation.

\subsubsection{Previously identified difficulties with physics content}\label{physics known}
Not surprisingly, there were several types of errors that were more connected to the physics than to the mathematics of these problems. Some of these errors have been previously discussed in the literature on student understanding of electromagnetism and some were associated with particular physics features of the problems.

Consistent with the quantitative analyses and with the findings of \citet{Scaife:2010,Scaife:2011}, there were two kinds of errors that were more associated with problems with \emph{Magnet} (or magnetic pole) representation for the magnetic field than with the \emph{Vector} representation. The first type of error was when the participant treated the direction of the magnetic field as pointing from South to North instead of from North to South (either a \emph{Direction} or \emph{Misinterpret} error).

The second error involved an inappropriate analogy to electric fields and forces (\emph{Electric Field}), although the analogy was rarely stated explicitly. In the \emph{Magnet} representation, several students stated that the magnetic force on a moving charge would be parallel to the field. For example, Ana answered that
\begin{quote}
``The force would be down\dots negative y, because it's pulled towards\dots the south and pushed away from north.''
\end{quote}

This kind of implicit analogy with electric field was also associated with problems that asked for the magnetic field at the center of a loop of current-carrying wire.
On these problems, eight of the participants responded at least once  like Danny did,
\begin{quote}
``\dots since it's in the center, they're all going to cancel out, so it just has to be\dots zero magnitude.''
\end{quote}

The fact that this type of error only appeared on problems with certain features suggests that there is something about the features that cued inappropriate electric field resources for these students. This interpretation is consistent with the bi-directional interference effects between electric and magnetic force noted by \citet{Scaife:2011}.

For the 2nd group of students  in this study who were interviewed later in the term, there was also some evidence of another interference effect, between magnetic field and magnetic force. For example, when Danny was trying to find the magnetic field produced by a moving charged particle, he tended to inappropriately try to apply a RHR designed to find the magnetic force on a moving charged particle. While not conclusive, this type of confusion implies that some students may not yet have differentiated the two different cross products (and associated right-hand rules) for these situations. 

The issues discussed here are consistent with previous literature on student understanding of electromagnetism and provide additional evidence of the prevalence of some of these student difficulties.

\subsubsection{Previously unidentified difficulties with physics content}\label{physics unknown}
There were two physics errors identified in this data set have not been previously discussed in the literature, but are consistent with anecdotal evidence from instructors. 

The first error, consistent with the quantitative analysis, was that participants would not always appropriately account for a negative charge. Sometimes this meant simply ignoring the sign of the charge, whereas other times, participants might ``double-count'' the negative charge. Of those that did not have \emph{Charge} errors, two (Fred and Danny) explicitly used a left-hand rule when there was a negative charge. Some students also expressed confusion about conventional versus electron current, which may be unique to Matter and Interactions curriculum \cite{MI3:em}.

The second error involved not appropriately accounting for the observation location on magnetic field questions. What makes this error even more interesting is that it appeared to be particularly associated with the `Current' RHR, where one points the thumb in the direction of the current and curls the fingers to the observation location. There were two participants (Ericka and Sebastien) who typically used this RHR and consistently did not curl their fingers to the appropriate location. Given that this particular RHR does not explicitly require identifying the observation location, it seems reasonable that those using this type of RHR might be more prone to this kind of error.

\subsection{Summary of qualitative analyses}
Expanding our definition of performance to include the methods that participants used and the types of errors that they made provides an additional window into how the contextual and representational features impact performance.

First, even students who state that the order of vectors in a cross product is important may switch this order. This happens particularly on problems requiring backward reasoning where you are looking for the first vector in the cross product. Using a \emph{Guess-and-Check} supplemental method for this kind of a problem may help to mitigate the difficulty. In addition, for students that do not recognize the non-commutativity of the cross product, errors may be mediated by other problem features, such as the physical awkwardness of the orientation.

Second, the quantitative analyses shows that students perform more poorly on problems with non-$xy$ plane orientations than on problems with $xy$ plane orientations. However, the qualitative analysis reveals that this trend may not be due solely to difficulties with mental rotations. Instead, the trend may be partly due to not correctly interpreting the symbols for vectors aligned with the $z$-axis ($\bigotimes$ and $\bigodot$).

Third, there is evidence that physically rotating the paper may help to mitigate difficulties associated with specific problem features associated with orientation, such as the physical awkwardness or an orientation that is not aligned with the vertical.

Fourth, the categorization of error types provides additional support for previous research on student difficulties with the \emph{Magnet} (or magnetic poles) representation, as well as for interference effects between electric and magnetic fields and between magnetic field and force.

Fifth, there are two errors associated with the physics contexts that have not previously been discussed in the literature, but are consistent with instructor experience: students struggle to appropriately account for negative charges (and electron currents) and the observation location when using RHRs. In particular, errors in accounting for the observation location seem to be associated with a particular right-hand rule for some students.

\section{Conclusions and Implications\label{conclusions}}
This study sought to answer the question:
\begin{quote}
How do contextual and representational features impact student performance on problems requiring the use of right-hand rules?
\end{quote}
Previous research was used to identify ten different problem features that were likely to impact performance (Fig.~\ref{problem features map}). Then  four different measures of performance and both quantitative and qualitative analyses were employed to address how those ten features impacted students use of RHRs. The quantitative analysis was primarily used to identify problem features that have the greatest impact on performance, as measured by response time and correctness. These results provided a focus for the qualitative analysis, which in turn revealed some of the nuances associated with the representational dependence of performance. Here we briefly summarize the findings from both analyses (Section~\ref{findings}) and discuss the implications for instruction (Section~\ref{implications})

\subsection{Review of findings\label{findings}}

\subsubsection{Kind of question: Context and Type}
Of all of the many features explored in this study, the type of reasoning required (working forward versus working backward) had the strongest impact on multiple aspects of performance: response time, correctness, and number of RHR errors made. All of these results show that working backward is cognitively more challenging than working forward, and in particular, finding the first vector of a cross product is the most difficult backward problem. The additional cognitive load of this type of problem may account for students treating the cross product as commutative even when they acknowledge the non-commutativity in other contexts. However, there are supplemental methods, such as \emph{Guess-and-Check} which may help to reduce this cognitive difficulty. 

The impact of context is less clear and requires further study, but there is support for previous research suggesting  interference effects between different physics contexts (\emph{i.e.}, electric vs. magnetic,  field vs. force, \emph{etc.}).

\subsubsection{Parallel Transport: Tails and Separation}
In the light of the literature on vector addition and subtraction, it was somewhat surprising to see that the features associated with parallel transport appeared to have little effect on performance. It may be that by this point in the physics sequence, parallel transport is no longer such a stumbling block to students. Alternatively, parallel transport may not be as necessary for using a right-hand rule as it is for graphical vector addition and subtraction.

\subsubsection{Orientation: Comfort, Plane, and Angle}
By treating the orientation of the vectors as three separate constructs instead of one, this study has resolved an apparent difference between the spatial cognition literature and \citet{Scaife:2010} on the impact of orientation. The quantitative analysis here showed that the degree of physical awkwardness, the plane of the vectors, and the angle of the vectors each have an impact on performance consistent with the spatial cognition literature. 

However, the qualitative analysis revealed that spatial cognition may not the only factor contributing to difficulties with orientations. The existence of \emph{Misinterpret} errors and their prevalence on problems with non-$xy$ orientations indicates that some of the difficulty with the plane of the vectors may stem from the symbols used to represent 3-dimensional vectors on a 2-dimensional page. In addition, one student, Andrea, was more likely to switch the order of the vectors on problems with physically awkward orientations, highlighting her belief in the commutativity of the cross product. This suggests that certain orientations may elicit errors that might otherwise not be apparent.

As with backward reasoning problems, there is evidence of a supplemental method reducing errors associated with orientation. Physically rotating the paper reduced the overall fraction of errors, as well as the gap between easy and challenging orientations. This trend was seen both with the different levels of physical awkwardness and with the different planes of the given vectors.

\subsubsection{Physics features: Charge, Shape, and Field}
This study identified several significant physics errors and related these errors to certain physics features. Each of these findings is consistent with previous literature and in some cases, 
 provides support for conventional wisdom that has been anecdotal until now.

This study confirmed the impact of the representation of magnetic field observed by \citet{Scaife:2010,Scaife:2011}, where participants perform more poorly on questions with a bar magnet (or magnetic pole) representation than on questions with a vector representation. The physics errors associated with this feature are also consistent with their findings. The first involved reversing the direction of the magnetic field and the second involved believing the force to be parallel to the field, likely using an inappropriate analogy to electric field. This confusion between electric and magnetic concepts, which has been explored in several previous studies~\cite{Maloney:2001,Guisasola:2002,Garza:2012}), was also seen on magnetic field problems where participants would state that the field ``points away from the positive and toward the negative'' or that the field at the center of a loop was zero.

The most significant physics errors in this data was not appropriately accounting for the sign of the charge. While most of these errors resulted from participants ignoring or neglecting the issue, some participants explicitly did not understand how the sign of the charge would impact the cross product and several students expressed confusion about conventional versus electron current.  

Finally, a previously un-explored issue in the correct application of RHRs is the effect of the observation location on the direction of the magnetic field. While not associated with a particular problem feature, observation location errors were one of the few instances where a specific type of right-hand rule was associated with a particular error. The majority of observation location errors were made when using the `Current' right-hand rule and for at least two of the participants (Ericka and Sebastien), this was a systemic issue.

\subsection{Implications for instruction\label{implications}}
For instructors who want to help students use RHRs effectively, this study has two primary implications:
\begin{enumerate}
\item Context and representation matter and it is important to attend to the features of a problem when thinking about instruction and assessment, and
\item There are other resources available that can be leveraged to alleviate some of the difficulties associated with certain problem features.
\end{enumerate}

The quantitative analysis showed several contextual and representational features that do, indeed, impact student performance and the qualitative analysis provided a more in-depth look at how they do so. For example, the problem in Fig.~\ref{order error} has several features that make it significantly more difficult than the problem in Fig.~\ref{misinterpret}: it is a physics problem that requires working backward and involves a negative charge. We want students to be able to perform appropriately on problems with different features, but we also need to be aware of the relative difficulty of the problems we assign and why we are assigning them. Providing a variety of orientations or scaffolding from more cognitively easy features to more difficult ones may help students to practice the skills that they need to be able to use RHRs appropriately in multiple contexts and with different representations. In addition, we should attend to possible interference effects from other concepts that students are learning, \emph{e.g.,} dot products vs. cross products, electric concepts vs. magnetic concepts, field vs. forces, \emph{etc.}

In addition to attending to the contextual and representational features of the problems we use and assign, we can and should leverage additional resources to provide students with additional tools that can help to ease some of the difficulties associated with some of these features. We can start with the tools that the participants in this study were using: \emph{Guess-and-Check}  as a way to reduce errors on \emph{Backward} problems and \emph{Rotation} as way to help with orientation issues. 

The role of Rotation in reducing errors associated with orientation suggests that manipulatives, 
such as those designed by \citet{VanDomelen:1999} and \citet{Nguyen:2003} could be effective in addressing orientation issues as well. Another option would be to leverage the spatial nature of the classroom through embodied learning activities. For example, one might have each student represents a point in space and use their bodies to show the direction of the field (or force) at their location due to a charge or current at some other location \cite{Kustusch:2014}.  This would not only emphasize that the field is different at different locations, but would also give the instructor formative assessment about each student's ability to use a RHR appropriately. One productive area for future inquiry would be to assess the effectiveness of such tools and activities. 

Also, given the rise of online resources, it is also important to explore how the impact of orientation would change for tasks on a computer, such with an online homework system (e.g., WebAssign or Mastering Physics) or with a computer simulation. The computer reference frame is more consistent with instructors' use of whiteboards, but exams are typically given on paper. The inability to rotate a 2-dimensional image on a computer screen as one can rotate a piece of paper could present challenges, but the use of rotatable computer simulations could provide additional tools for students to explore the 3-dimensional nature of cross products.

Finally, the difficulty that some participants had in appropriately accounting for the observation location is an issue that has not previously been discussed as such in the literature and has potentially far reaching implications. For physics majors, much of upper-division electromagnetism involves exploring and assessing patterns of fields in space where understanding the role of the separation vector ($\vec{r}-\vec{r}'$) is vital. Thus, it is important to explore to what extent the issue identified here is due to an inappropriate application of a RHR and to what extent this is a more foundational conceptual difficulty with the role of the observation location.

This study has demonstrated that when it comes to using a right-hand rule appropriately, the contextual and representational features make a difference. If we want to help students to use right-hand rules effectively, we may need to leverage other resources to help to ease the cognitive difficulties associated with these representations.

\begin{acknowledgments}
Grateful thanks  to all of the North Carolina State University PER group, especially Bin Xiao, who provided interrater reliability, and  Bob Beichner, who mentored me through this project. Funding for this project was provided by the NCSU STEM Initiative and NSF award DUE--0618504. 

Special thanks to Amy Robertson for providing invaluable feedback on initial drafts of this manuscript.
\end{acknowledgments}


\appendix

\begin{table*}
\caption{Definitions for all methods (see \citet{Kustusch:dissertation} for more detailed code definitions and examples)}
\label{method code definitions full}
\begin{ruledtabular}
\begin{tabular}{lp{5.25in}}
Code Name& Code Definition\\
\hline
\\
RIGHT-HAND RULE (RHR) & Participant uses a physical technique to find the direction of the cross product.\\
\\
&\begin{tabular}{l p{4.5in}}
\multicolumn{2}{l}{Sub-code Definition}\\
\hline
\\
HAND & Specifies whether participant uses RIGHT or LEFT hand\\
\\
FRAME	
& Specifies whether the RHR is performed in the plane of the table (TABLE), where $+ \hat{x}$ is to the right, $+ \hat{y}$ is away from the participant, and $+ \hat{z}$ vertically upward, or in a different (OTHER) frame, e.g.,  $+ \hat{z}$ is toward participant instead of vertically upward.\\
\\
TYPE	 &Specifies the type of right-hand rule.\\
\\
& STANDARD: Fingers point in the direction of the first vector ($\vec{A}$), curl in the direction of the second vector ($\vec{B}$) and the thumb points in the direction of $\vec{A}\times \vec{B}$.\\	
\\
& CURRENT: Thumb points in the direction of the conventional current and the fingers curl in the direction of the magnetic field, such that at a given observation location, the fingers point in the direction of the magnetic field at that location.\\
\\
& LOOP: Fingers curl around a loop of wire and thumb gives the direction of the magnetic field at the center of the loop.\\
\\
& 3 FINGERS: The index finger points in the direction of the first vector ($\vec{A}$), the middle finger points in the direction of the second vector ($\vec{B}$) and the thumb points in the direction of $\vec{A}\times \vec{B}$.\\	
\\
& KNUCKLES: Knuckles point in the direction of the first vector ($\vec{A}$), fingers in the direction of the second vector ($\vec{B}$) and thumb points in the direction of $\vec{A}\times \vec{B}$.\\	
\\
& PALM: Thumb points in the direction of the first vector ($\vec{A}$), fingers in the direction of the second vector ($\vec{B}$) and palm points in the direction of $\vec{A}\times \vec{B}$.\\	
\end{tabular}\\
\\

GUESS \& CHECK & Participant explicitly ``guesses'' an answer or makes an assumption (e.g.,  a particle has positive charge) and uses another method to determine whether the assumption is consistent with the given information.\\
\\
ROTATION & The given frame of reference is rotated by more than $\approx15^\circ$ using either the paper or a diagram.\\
\\
DIAGRAM & Participant draws a diagram. \\
\\
MULTIPLICATION & Participant uses either commutative or non-commutative multiplication rules outside of an explicit calculation (e.g., two negatives make a positive). \\
\\
PARALLEL TRANSPORT & Participant explicitly mentions (or gestures or draws) moving the vectors.\\
\\
ORTHOGONALITY & Participant restricts the solution to one axis before using a right-hand rule (e.g., since given vectors are along $x$ and $y$ axes, the cross product must be $\pm \hat{z}$).\\
\\
PHYSICS KNOWLEDGE & Participant explicitly mentions physics not present in the problem statement or uses physics information from the problem statement to reason about the problem. \\
\\
ADDITION & Participant uses methods consistent with vector addition or subtraction.\\
\\
MATRIX & Participant makes use of matrices and determinants in finding the direction of the cross product. \\
\\
UNCLEAR &Not enough evidence to determine what method(s) participant used to solve the problem.
\end{tabular}
\end{ruledtabular}
\end{table*}

\begin{table*}
\caption{Definitions for all errors types (see \citet{Kustusch:dissertation} for more detailed code definitions and examples). Unless otherwise specified, the definition corresponds to an ERROR sub-code, where sub-codes of CORRECTED or NO ERROR are also possibilities. For CHARGE and ELECTRIC FIELD, explicit and implicit errors were categorized separately.}
\label{error code definitions full}
\begin{ruledtabular}
\begin{tabular}{p{1.5in}p{5.25in}}
Code Name& Code Definition\\
\hline
\\
ADDITION & Participant uses vector addition or subtraction method (e.g., connects tip to tail or tip to tip).\\
\\
MATRIX &Participant makes at least one error with the use of matrices that contributes to final response.\\
\\
RIGHT-HAND RULE &\\
&\begin{tabular}{l p{4in}}
\multicolumn{2}{l}{Sub-code Definition}\\
\hline
\\
NO ERROR& Participant makes no errors with this RHR regardless of whether the final response includes an error based on another method. \\
\\
INAPPROPRIATE	& Participant attempts to make use of a right-hand rule that is inappropriate for the given situation. If any other error (e.g. Direction or Order) is made, this code takes precedence. \\
\\
DIRECTION	 &Participant inappropriately reverses the direction of one of the given vectors by explicitly stating correct direction and using the wrong direction or by using the wrong direction without stating it. \\
\\
ORDER & Participant does the right hand rule in such a way that the order of the vectors is reversed and they are performing $\vec{B} \times \vec{A}$ instead of $\vec{A} \times \vec{B}$. \\	
\\
RHR OTHER &Participant uses a right-hand rule such that it is unclear what error(s) they are making or the error does not fall into one of the other right-hand rule categories or they make multiple errors. \\
\end{tabular}\\
\\
OBSERVATION LOCATION& Participant does not appropriately account for the observation location, either implicitly or explicitly (categorized together). \\
\\
CHARGE &Participant does not appropriately account for the sign of the charge, either implicitly or explicitly (categorized separately). \\
\\
ELECTRIC FIELD& Participant explicitly reasons from, provides an answer consistent with, an invalid or inappropriate analogy to electric field (implicit and explicit errors categorized separately). \\
\\
 PHYSICS OTHER & Participant explicitly invokes incorrect or inappropriate physics not covered in the other ``Physics'' categories or uses the same wrong physics on question(s) immediately following an explicit error without restating the physics. \\
\\
MISINTERPRET& Participant misinterprets the diagram or misreads the problems statement in a way that contributes to the final response. \\
\\
MATCH & Participant gives a verbal solution that does not match the gestural (or diagrammatic) response. \\
\\
OTHER& There is at least one error that contributes to the final response which is not due to a primary method, physics, misinterpretation or a gestural mismatch. \\
\end{tabular}
\end{ruledtabular}
\end{table*}

\bibliography{RHR}

\end{document}